\documentclass[aps,twocolumn,superscriptaddress,groupedaddress]{revtex4}

\usepackage{graphicx}  % needed for figures
\usepackage{dcolumn}   % needed for some tables
\usepackage{bm}        % for math
\usepackage{amssymb}   % for math
\usepackage{amsmath}
\usepackage{epstopdf}
\begin{document}

\title{Polariton lasing and energy-degenerate parametric scattering \\ in non-resonantly driven coupled planar microcavities}

\author{K.~Sawicki} \affiliation{Institute of Experimental Physics, Faculty of Physics, University of Warsaw, Pasteura 5, PL-02-093 Warsaw, Poland}
\author{T.~J.~Sturges} \affiliation{Institute of Theoretical Physics, Faculty of Physics, University of Warsaw, Pasteura 5, PL-02-093 Warsaw, Poland}
\author{M.~\'Sciesiek} \affiliation{Institute of Experimental Physics, Faculty of Physics, University of Warsaw, Pasteura 5, PL-02-093 Warsaw, Poland}
\author{T.~Kazimierczuk} \affiliation{Institute of Experimental Physics, Faculty of Physics, University of Warsaw, Pasteura 5, PL-02-093 Warsaw, Poland}
\author{K.~Sobczak} \affiliation{Biological and Chemical Research Centre, University of Warsaw, \.{Z}wirki i Wigury 101, 02-089 Warsaw, Poland}
\author{A.~Golnik} \affiliation{Institute of Experimental Physics, Faculty of Physics, University of Warsaw, Pasteura 5, PL-02-093 Warsaw, Poland}
\author{W.~Pacuski} \affiliation{Institute of Experimental Physics, Faculty of Physics, University of Warsaw, Pasteura 5, PL-02-093 Warsaw, Poland}
\author{J.~Suf\mbox{}fczy\'nski} \affiliation{Institute of Experimental Physics, Faculty of Physics, University of Warsaw, Pasteura 5, PL-02-093 Warsaw, Poland}
\vskip 0.25cm

\begin{abstract}
Multi-level exciton-polariton systems offer an attractive platform for studies of non-linear optical phenomena. However, studies of such consequential non-linear phenomena as polariton condensation and lasing in planar microcavities have so far been limited to two-level systems, where the condensation takes place in the lowest attainable state. Here, we report non-equilibrium Bose-Einstein condensation of exciton-polaritons and low threshold, dual-wavelength polariton lasing in vertically coupled, double planar microcavities. Moreover, we find that the presence of the non-resonantly driven condensate triggers interbranch exciton-polariton transfer in the form of energy-degenerate parametric scattering. Such an effect has so far been observed only under excitation that is strictly resonant in terms of the energy and incidence angle. We describe theoretically our time-integrated and time-resolved photoluminescence investigations by a set of rate equations involving an open-dissipative Gross-Pitaevskii equation. Our platform's inherent tunability is promising for construction of planar lattices, enabling three-dimensional polariton hopping and realization of photonic devices, such as two-qubit polariton-based logic gates.
\end{abstract}

\maketitle

\section{Introduction}

Strong coupling between excitons and an optical mode leads to the formation of a hybrid quasiparticle called a polariton.\cite{Weisbuch:PRL1992} The polaritons can be accessed and manipulated by their light component while the matter, excitonic component is responsible for the non-linear nature of the light-matter coupling and gives the polaritons the ability to interact.

In particular, the excitonic component is at the origin of polariton scattering processes, either of polariton-polariton type or related to interaction with the environment.\cite{Ciuti:PRB2000, Krizhanovskii:PRB2007, Amo:NatPhys2009} In semiconductor microcavities non-degenerate parametric scattering becomes efficient when a non-resonantly created high polariton density accumulates at the {\it bottleneck} region of the dispersion curve or when the resonant excitation is adjusted to the so-called {\it magic angle} of incidence.\cite{Savvidis:PRL2000} The bosonic nature of polaritons enables their massive occupation of a single quantum state. Stimulated parametric scattering to the final state leads to such fascinating and consequential effects as Bose-Einstein condensation of polaritons,\cite{Deng:Science2002,Kasprzak:Nature2006} achievable at ultra-low excitation power and a higher temperature than in cold-atomic systems.\cite{Anderson:Science1995} Photons emitted in radiative recombination of the condensate inherit its coherence and phase. By analogy to the conventional laser, the emission arising from the condensate is described by the term {\it polariton lasing}.\cite{Imamoglu:PRA1996,Deng:PNAS2003,Christopoulos:PRL2007,Bajoni:PRL2008}

Studies of polariton condensation and related effects such as spontaneous coherence, lasing, or superfluidity have increasingly become a focus of condensed matter and optics research in recent years. However, they have so far been mostly limited to single microcavities, either planar\cite{Deng:PNAS2003,Kasprzak:Nature2006, Amo:NatPhys2009} or microstructured,\cite{Galbiati:PRL2012,Grosso:PRB2014,Klein:APL2015,Rodriguez:NatComm2016, Su:NanoLett2017, Su:NatPhys2020,Zambon:PRA2020} and were related to intrabranch, non-degenerate polariton parametric scattering.

Extension of studies of polariton-related non-linear phenomena to such effects as energy or momentum degenerate polariton parametric scattering,\cite{Ardizzone:SciRep2013,Diederichs:Nature2006} localized to delocalized phase transitions of a photon,\cite{Hartmann:NaturePhys2006} non-reciprocity based optical isolation,\cite{Armitage:PRB1998,Chang:NaturePhot2014} parity-time symmetry breaking,\cite{Hamel:NatPhot2015} or the generation of non-classical states of light\cite{Gerace:PRA2014,Portolan:NJofPhys2014} requires implementation of a multiple-level polariton system. Such a system has so far been realized in a set of vertically coupled planar microcavities\cite{Stanley:APL1994,Diederichs:Nature2006} or a semiconductor microrod.\cite{Xie:PRL2012,Trichet:NJP2012,Xu:APL2014,Zhang:Nanoscale2018} Offering a high degree of tunability, vertically coupled microcavities have been studied mostly in the linear regime,\cite{Stanley:APL1994,Armitage:PRB1998,Stelitano:APL2009, Liu:PRB2015,Sciesiek:CGD2017,Jayaprakash:LSA2019,Sciesiek:CommunMat2020} while the non-linear regime was addressed in the exclusive context of resonantly driven parametric scattering.\cite{Diederichs:Nature2006,Ardizzone:SciRep2013} In particular, such fundamental effects as polariton condensation and lasing, or parametric polariton scattering under non-resonant excitation, have so far not been observed and studied in a vertically coupled, planar microcavity system.

\begin{center}
\begin{figure*}
\includegraphics[width=\textwidth]{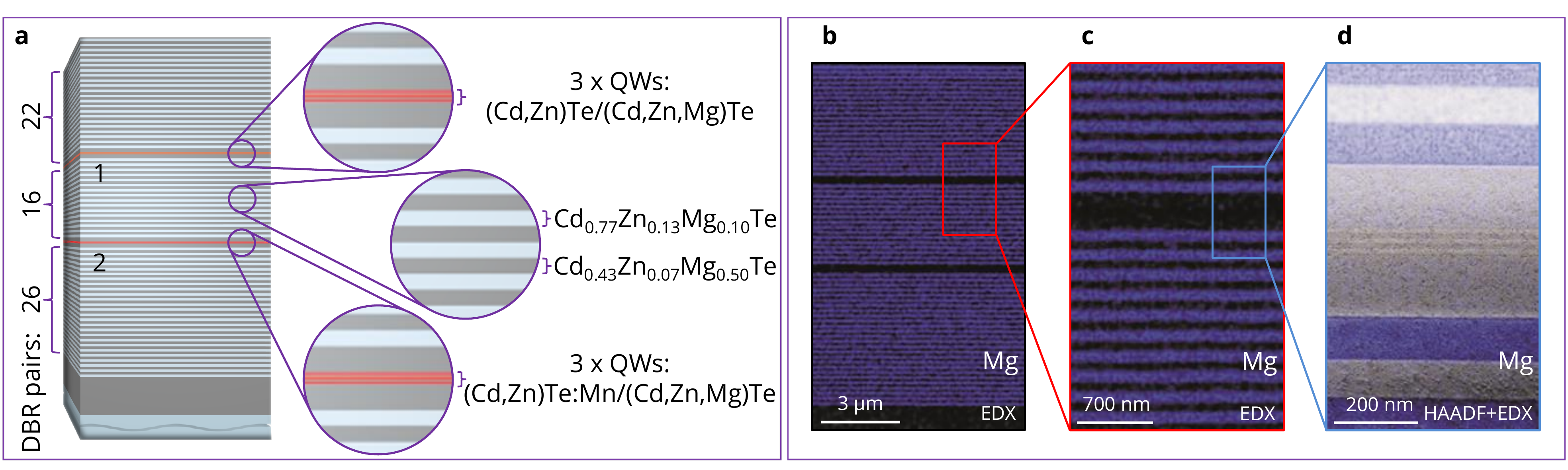}
\caption{\label{fig:scheme} {\bf a} Sketch of a vertically coupled, double microcavity structure. The top and the bottom microcavities contain a set of three (Cd,Zn)Te or (Cd,Zn)Te:Mn QWs, respectively. {\bf b, c} Cross-sectional energy-dispersive X-ray spectroscopy (EDX) images of the sample showing the spatial distribution of magnesium, the content of which controls the energy bandgap and refractive index of the layers.\cite{Andre:JAP1997} {\bf d} A high angle-annular dark-field scanning transmission electron microscope image with EDX compositional mapping of magnesium in the region of the microcavity with three QWs visible.}
\end{figure*}
\end{center}

In this letter we implement vertically coupled double planar microcavities, each embedding quantum wells (QW), to observe non-linear polariton phenomena under non-resonant excitation. In two-level polariton systems the fast relaxation from the upper to the lower branch precludes polariton condensation in the upper polariton state. Here, we obtain polariton condensation not only in the lowest, but also in the upper polariton branch. Emission dynamics measurements reveal that the two condensates, differing in energy and excitonic content, do not coexist simultaneously at the same point in the structure. We introduce a theoretical model, assuming the existence of two polariton reservoirs, an active and an inactive one, which accurately describes both the time-integrated and time-resolved data. In particular, the model confirms different formation and recombination dynamics of the two condensates and indicates that the interplay between the relaxation and loss kinetics governs the condensation process. Moreover, we show that the polariton condensate created via non-resonant excitation in the upper polariton branch triggers energy degenerate parametric scattering from the bottom of the upper branch to a high wave vector within the lowest polariton branch. In this way, the condensate acts equivalently to a pump laser tuned to resonance with the bottom of the polariton branch.

\section{Sample}

We investigate a molecular beam epitaxy grown sample with two $3\lambda/2$ Cd$_{0.77}$Zn$_{0.13}$Mg$_{0.10}$Te microcavities coupled via a semi-transparent Distributed Bragg Reflector (DBR) (see Figure~\ref{fig:scheme}~{\bf a}). The DBRs are made of alternating Cd$_{0.77}$Zn$_{0.13}$Mg$_{0.10}$Te and Cd$_{0.43}$Zn$_{0.07}$Mg$_{0.50}$Te layers lattice-matched to MgTe.\cite{Rousset:JCG2013, Pacuski:CGD2017, Sciesiek:CommunMat2020} Three quantum wells are placed at the anti-node of the electric field of each of the microcavities: 12~nm wide (Cd,Zn)Te:Mn QWs in the lower and 10~nm wide (Cd,Zn)Te QWs in the top microcavity. The microcavities and the DBRs are wedge-like, meaning that the absolute thickness of the layers of the sample changes when varying position across the sample's surface, whereas the ratio of the thicknesses of all the layers stays constant. This allows us continuously to tune the energy of the optical modes in a 100~meV wide range by adjusting the position on the sample surface while keeping the cavity-cavity coupling strength unaltered.

The results of a characterization of the structure by scanning transmission electron microscopy (STEM) is shown in Figures~\ref{fig:scheme}~{\bf b-d}. Consecutive close-up views of the microcavity structure present sharp interfaces of the DBR and QW layers, testifying to the high structural quality of the sample.

\section{Experiment}

The sample is placed in a liquid-helium flow or pumped-helium cryostat and cooled down to 8~K or 1.5~K, respectively. The emission is pumped non-resonantly at E$_{{\rm exc}}$ = 1.72~eV ($\lambda_{{\rm exc}}$ = 720~nm) at normal incidence using a pulsed Ti:sapphire laser operating in femtosecond mode. The excitation power is adjusted using neutral density filters. To avoid saturation of the detectors by stray laser light, detection is linearly cross-polarized relative to the excitation. We use a CCD or a streak camera combined with a grating spectrometer (1200 grooves/mm) as a detector for time-integrated or time-resolved measurements, respectively. The laser beam is focused onto the sample surface to a spot of 1-2~$\mu$m diameter, and the signal is collected with a microscope objective (NA~=~0.7) or an aspheric lens (f~=~3.1~mm, NA~=~0.68) for time-integrated or time-resolved measurements, respectively. By adjusting a set of lenses, we switch between acquiring angle-integrated signal and Fourier space imaging. The in-plane photon momentum wave vector is recorded over a range of values up to 4.2~$\mu m^{-1}$. Reflectivity spatial mapping of the structure is performed using a halogen lamp as the light source, with a lens (focal length of 500~mm) shifted in the sample plane (step of 0.06~mm). The size of the light spot on the sample surface is 0.1~mm in this case.

\section{Results}

\begin{center}
\begin{figure}
\includegraphics[width=0.8\linewidth]{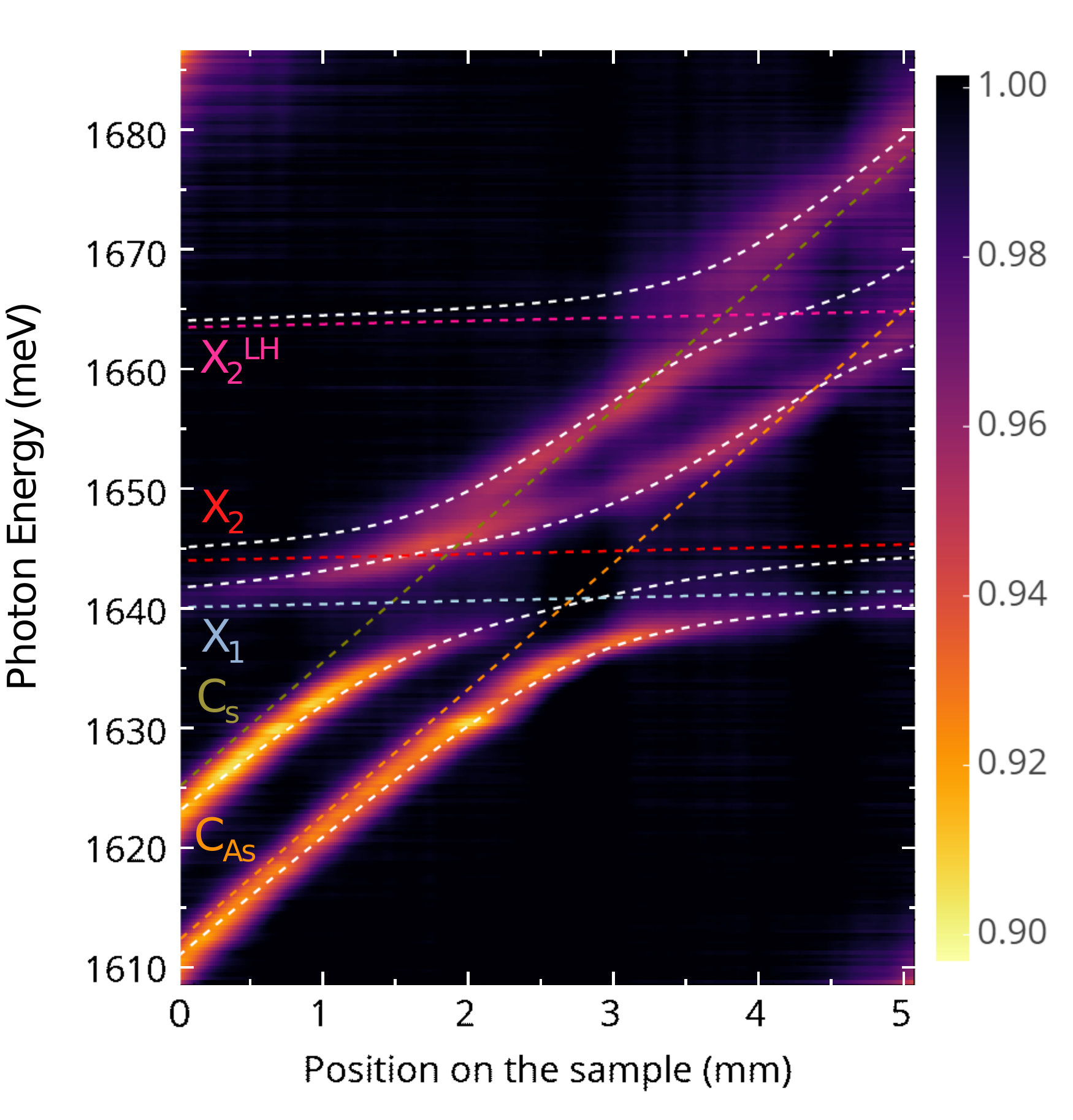}
\caption{\label{fig:Ref_map} Reflectivity spectra of the vertically coupled, double microcavities, registered as a function of the detuning between the microcavities' coupled modes and QW excitons. The lines, as indicated by the labels, describe the bare levels of the heavy hole excitons X$_{\text{1}}$ and X$_{\text{2}}$ in the (Cd,Zn,Te) QWs and (Cd,Zn,Te):Mn QWs, respectively, the light-hole exciton X$_{\text{2}}^{LH}$ in the (Cd,Zn,Te):Mn QWs, as well as the asymmetric $C_\mathrm{AS}$ and symmetric $C_\mathrm{S}$ modes of the coupled microcavities. White, unlabelled lines represent calculated polariton states.}
\end{figure}
\end{center}

Figure~\ref{fig:Ref_map} shows the reflectivity spectra recorded for consecutive positions on the sample along the microcavities' thickness gradient. The coupled, delocalized spatially optical modes of the structure C$_S$ and C$_{AS}$ shift in energy with the variation of the position on the sample, entering in resonance with consecutive excitonic transitions. The heavy-hole excitons, X$_{\text{1}}$ in the (Cd,Zn,Te) QWs at 1640~meV and X$_{\text{2}}$ in the (Cd,Zn,Te):Mn QWs at 1644~meV, as well as the light-hole exciton X$_{\text{2}}^{LH}$ in the (Cd,Zn,Te):Mn QWs at around 1663.5~meV are seen. The energies of the polariton levels are simulated within the framework of a coupled oscillator model (see Supplementary Note 1), as indicated by the white lines. The presence of strong coupling conditions is manifested in Figure~\ref{fig:Ref_map} by the anticrossing of the polaritonic resonances.\cite{Armitage:PRB1998,Richter:APL2015} Matching of the calculated energies to the experimental ones allows us to determine the coupling constant $\Omega_\mathrm{1}$~=~12~meV~$\pm$~0.3~meV between the mode of the top microcavity and X$_{\text{1}}$, as well as $\Omega_\mathrm{2}$ = 10.6~meV~$\pm$~0.3~meV or $\Omega_\mathrm{2}^{LH}$ = 9.8 meV~$\pm$~0.4~meV between the mode of the bottom microcavity and X$_{\text{2}}$ or X$_{\text{2}}^{LH}$, respectively. Modes of the top and bottom microcavity are coupled with a constant $\kappa_\mathrm{1,2}$ = 12.8~meV~$\pm$~0.2~meV, governing the C$_S$ and C$_{AS}$ energy separation. One should note here that, unlike in the case of the two-level polaritonic system, the map in Figure~\ref{fig:Ref_map} shows a crossing of coupled (polariton) and uncoupled (optical modes and exciton states) levels. Such behaviour is a property of any system involving a number of polariton levels higher than two.

\begin{figure}
\begin{center}
\includegraphics[width=\linewidth]{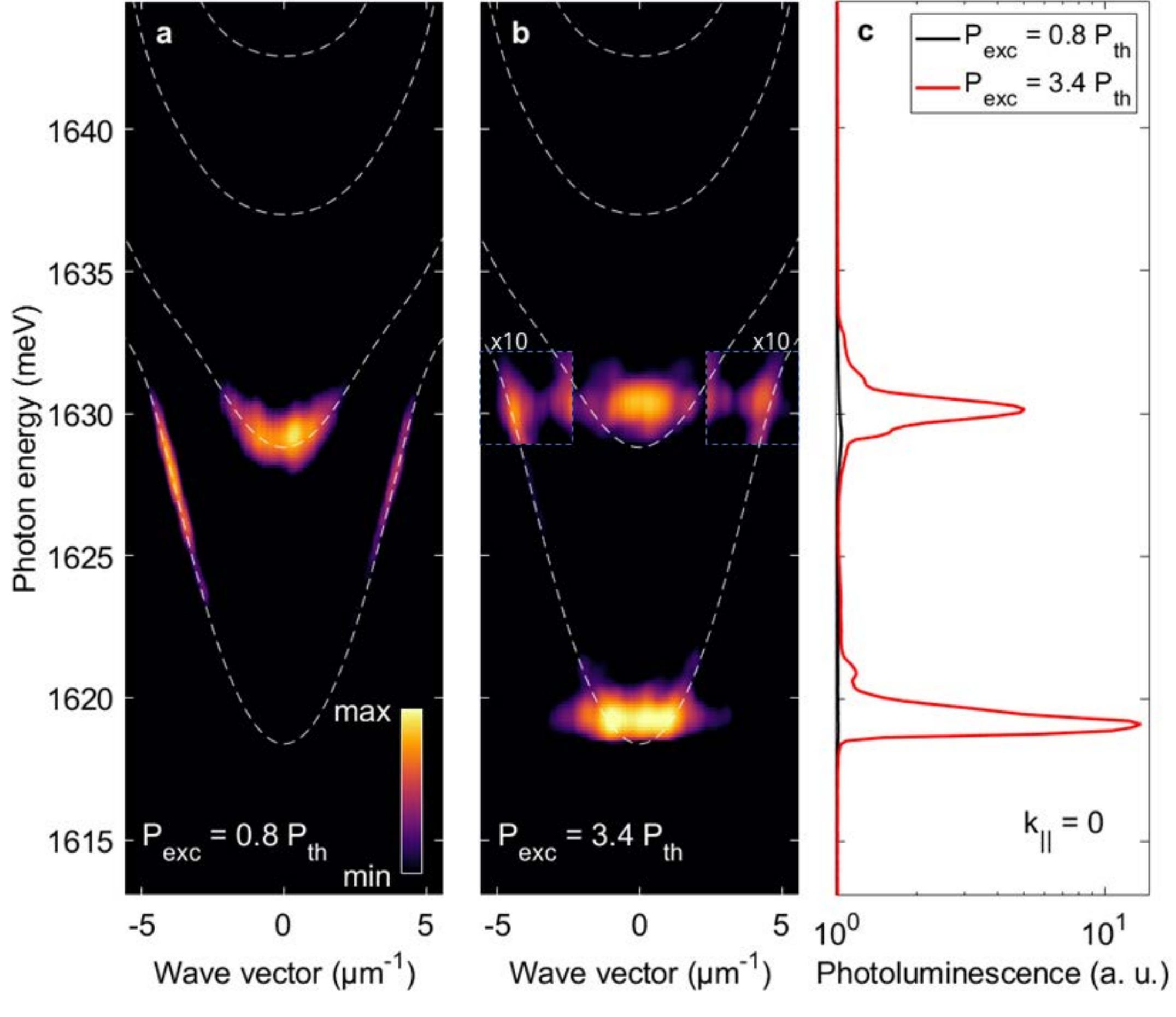}
\end{center}
\caption{\label{fig:below_above} Emission spectrum of vertically coupled, double microcavities resolved in in-plane photon momentum below (P$_{exc}$~=~0.8~P$_{th}$) and above (P$_{exc}$ = 3.4 P$_{th}$) the excitation density corresponding to the polariton lasing threshold at P$_{th}$~=~25~kW~cm$^{-2}$. White dashed lines represent calculated polariton levels. The indicated portion of the plot is multiplied by a factor of 10.}
\end{figure}

The non-parabolic dispersion of polariton branches observed in non-resonantly pumped photoluminescence (see Figure~\ref{fig:below_above}) recorded at a position on the sample of 0.7~mm, further confirms the strong light-matter coupling regime in the studied structures. For excitation density P$_{exc}$ = 0.8~P$_{th}$ (Figure~\ref{fig:below_above}~{\bf a}), the emission from the bottom of the upper polariton branch at 1619~meV dominates the spectrum. The much weaker emission from the lower branch is stretched along the dispersion curve in its {\it bottleneck} region. When the excitation density is increased up to P$_{exc}$~=~3.4~P$_{th}$ (Figure~\ref{fig:below_above}~{\bf b}), a strong emission limited to the close vicinity of k$_{\parallel}$~=~0 is observed from both the upper and the lower branch. Due to the relatively small size of the excitation spot and the resulting high density reservoir of photo-created carriers, the polaritons are partially ejected out of the bottom of the dispersion curves. A qualitative change in the spectrum shape at k$_{\parallel}$~=~0 with increasing excitation power density is depicted in Figure~\ref{fig:below_above}~{\bf c}.

\begin{center}
\begin{figure}
\includegraphics[width=0.8\linewidth]{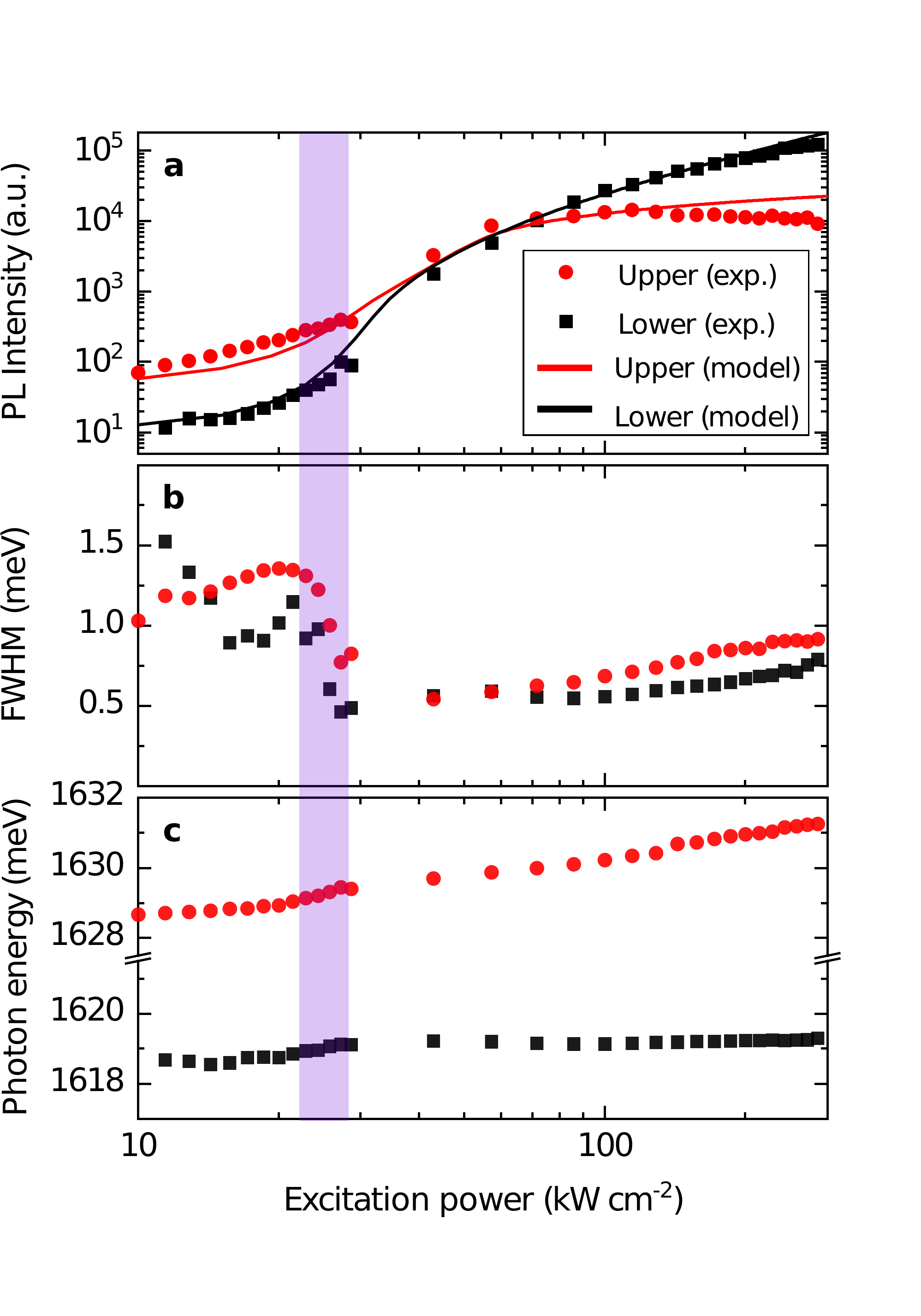}
\caption{\label{fig:lasing} Emission properties of the two lowest polariton branches at k$_{\parallel}$ = 0 as a function of the excitation density (points). Polariton condensation and lasing occur for both levels at approximately the same threshold, as revealed by {\bf a} a non-linear increase in intensity, {\bf b} narrowing and {\bf c} energy blueshift of the emission. Solid lines in {\bf a} represent the result of the model calculation (see text for details).}
\end{figure}
\end{center}

In order to trace in detail the impact of the excitation density on the emission properties of the studied structure, a systematic measurement of input-output dependence is performed with a focus on the emission at k$_{\parallel}$~=~0 from the two lowest polariton levels. The emission intensity of both levels increases non-linearly by more than three orders of magnitude across the threshold at around P$_{th}$ = 25~kW~cm$^{-2}$ (see Figure~\ref{fig:lasing}~{\bf a}). Crossing the threshold is assisted by the narrowing (Figure~\ref{fig:lasing}~{\bf b}) and blueshift (Figure~\ref{fig:lasing}~{\bf c}) of the emission in the case of both levels. The higher exciton content in the polariton, the larger magnitude of polariton-polariton interactions, hence the blueshift of the more excitonic upper level is larger than that of the more photonic lower level. With further increase of the power density the emission from the upper level saturates, while the signal from the lower level increases linearly. Such properties of the emission allow us to attribute the massive occupation of the bottom of the polariton branches shown in Figure~\ref{fig:below_above} upon crossing the threshold P$_{th}$ to the effect of polariton condensation.\cite{Deng:PNAS2003,Kasprzak:Nature2006,Sawicki:CommPhys2019}

In the studied four-level system polariton condensation and lasing occur at the two lowest levels, in contrast to the typical case of a two-level polariton system, where condensation takes place in the ground state, that is at the bottom of the lowest branch. Moreover, the condensation threshold for both the lower and upper polariton levels is comparable, unlike what was previously observed in ZnO microwire-based multimode systems.\cite{Liao:APE2019} As Figure~\ref{fig:lasing}~{\bf a} shows, tuning of excitation power allows for a steering of the relative emission intensity of the upper and lower levels. In particular, switching of polariton lasing from the lower to the upper branch is achievable.

In order to describe the dependence of the emission intensity on the excitation power a coupled rate equations model involving an open-dissipative Gross-Pitaevskii equation (see  Supplementary Note~2), inspired by Refs.~[\cite{Wouters:PRB2008,Lagoudakis:PRL2011,Winkler:PRB2016}], was implemented. The model assumes that electron-hole pairs created by the non-resonant, pulsed excitation with equal efficiency in both microcavities\cite{Sciesiek:CommunMat2020} accumulate in an inactive reservoir. From the inactive reservoir they either decay or relax to an active reservoir in the high energy and momentum region of the lower and upper polariton dispersions.\cite{Lagoudakis:PRL2010,Lagoudakis:PRL2011,Veit:PRB2012,Anton:PRB2013} The polaritons then relax towards the bottom of a given branch. When the polariton density is high enough, stimulated scattering to the minimum of the polariton branch accelerates the relaxation and, eventually, induces the condensation of polaritons in the bottom of the branch. The model also includes a direct transfer from the upper to the lower polariton branch at k$_{\parallel}$~=~0. Very good agreement between the experimental data (points) and the calculation (solid lines) is seen in Figure~\ref{fig:lasing}~{\bf a}. A detailed description of the model along with the values of the parameters ensuring agreement with the experimental data is provided in Supplementary Note~2.

\begin{figure}
\includegraphics[width=\linewidth]{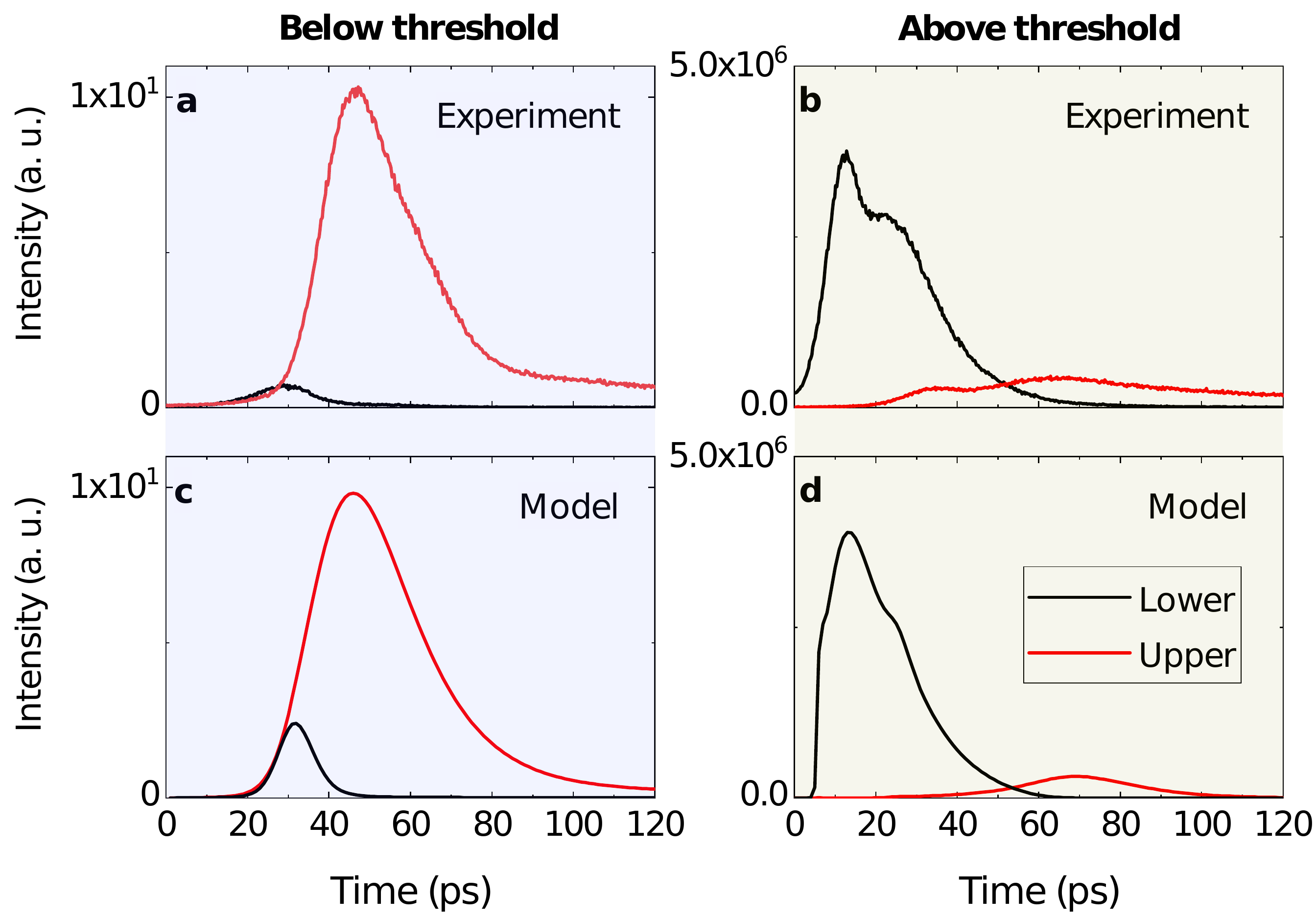}
\caption{\label{fig:timemodel} {\bf a, b} Experimental and {\bf c, d} calculated emission dynamics of the upper
(red line) and the lower (black line) polariton level at k$_{\parallel}$ = 0. The spectra shown in {\bf a} and {\bf c}) are acquired and calculated for an excitation power density of 0.8 P$_{th}$, while for the spectra in panels {\bf b} and {\bf d}) the excitation power density is 6.4 P$_{th}$.}
\end{figure}

To answer the question of how two condensates of different energy may coexist at the same point of the sample, we perform emission dynamics measurements. The temporal cross-sections of the spectra obtained under non-resonant excitation at the energy of the upper and lower polariton levels extracted from streak camera images are shown in Figure~\ref{fig:timemodel}. For excitation density below P$_{th}$ the population of the upper level (that with a higher excitonic content) builds up first and quickly decays. Once it vanishes, the population of the lower level (with a higher photonic content) builds up. The same behaviour is observed for excitation density above P$_{th}$, however, the intensity of the lower level is much higher than that of the upper level, and an oscillatory character of the decay curve is observed.

The experimental data presented in Figure~\ref{fig:timemodel}~{\bf a} and {\bf b} are very well described by the calculated time transients shown respectively in Figure~\ref{fig:timemodel}~{\bf c} and {\bf d}, both in terms of temporal dynamics and intensity. It should be stressed that the same values of the parameters are used in the description of both time-integrated and time-resolved photoluminescence presented in Figure~\ref{fig:lasing}~{\bf a} and Figure~\ref{fig:timemodel}, respectively. We only allow for a scaling of the excitation power and intensity axes by an arbitrary factor common to all the calculated curves to account for the unknown excitation and detection efficiency of the setup.

In view of our simulations, the relative temporal order of the emission from the bottom of the lower and upper branch is governed by the relationship between the rates of transfer from the active reservoir to these levels, independently of the excitation power. Namely, emission from the lower level prior to emission from the upper level is induced by a higher transfer rate from the active reservoir to the lower level than to the upper level. The relative emission intensity ratio of the lower to the upper level is affected by both the transfer rates and the decay times of the polariton population of these levels, \emph{i.e.}, below the condensation threshold, the shorter the lifetime the stronger the emission of a given level for a reasonably wide range of values and ratios of the transfer rates. Above P$_{th}$, the role of the transfer rates becomes dominant. In turn, the saturation of the emission intensity from the upper level above P$_{th}$ results from the efficient transfer of the condensate from the upper to the lower level. A possible mechanism of the interbranch transfer at k$_{\parallel}$ = 0 is energy relaxation assisted by the emission of acoustic phonons or momentum degenerate polariton scattering.\cite{Diederichs:Nature2006} Oscillations in the emission decay are a consequence of the presence of the inactive reservoir and competition in the polariton population built-up in the upper and lower polariton branches.

Upon the formation of the polariton condensate in the upper level a signal emerges at discrete points at high k$_{\parallel}$ of the lowest polariton branch, as seen in the momentum-resolved emission visible in Figure~\ref{fig:below_above}~{\bf b}. These discrete points are distributed symmetrically around k$_{\parallel}$~=~0, at k$_{l}$~=~-4.1~$\mu$m$^{-1}$ and k$_{r}$~=~4.1~$\mu$m$^{-1}$, at an energy of 1630.5~meV, that is the energy of the condensate formed in the upper polariton level. We interpret the presence of such points as a manifestation of energy degenerate parametric scattering from the upper to the lower polariton branch, \emph{i.e.} the annihilation of two polaritons with a wave vector k$_{\parallel}$ = 0 and creation of two polaritons with wave vectors {\bf k}$_l$ and {\bf k}$_r$. Such a process fulfils the condition {\bf k}$_r$ + {\bf k}$_l$ = 0 while keeping the polariton energy unchanged. The scattering is facilitated by negative detuning conditions and the resulting relatively low exciton content in the polaritons, which diminishes the absorption. Point-like emission at k$_{l}$ and k$_{r}$ appears only for excitation density above P$_{th}$ and is not observed below P$_{th}$, as further described in Supplementary Note 3. This indicates that the presence of a polariton condensate is a prerequisite for the observed energy degenerate parametric scattering of polaritons. In the previous studies on planar coupled microcavities energy degenerate parametric scattering was achieved by tuning the excitation to a strict resonance with the bottom of the upper polariton branch.\cite{Ardizzone:PRB2012,Ardizzone:SciRep2013,Lecomte:PRB2014} Our work shows, in contrast, that a non-resonantly driven polariton condensate provides an alternative to the previously used resonant pumping. The use of a non-resonant excitation for preparation of the condensate enables a clean observation of the scattering effects thanks to the spectral separation of the excitation and the measured signal. It is also promising for implementation of robust, polariton-based entangled photon sources.

\section{Discussion}

Studies leading to an understanding of non-linear phenomena including condensation and interbranch polariton scattering in multi-level polariton systems are essential from the point of view of the development of such domains as topological photonics\cite{Jacqmin:PRL2014,Kokhanchik:arxive2020} or polaritronics.\cite{Gao:PRB2012} A high degree of tunability of the system should allow the study of various channels of energy- and/or momentum degenerate polariton pair scattering under non-resonant pumping.\cite{Dasbach:PRB2002, Ciuti:PRB2004} Manganese doping of the QWs in one of the microcavities opens perspectives for a quite natural extension of the present study to spin-related phenomena, such as conservation of spin in energy degenerate parametric scattering and control of the polariton condensation in the multi-level system using a magnetic field. In turn, a microstructurization of the structure would open exciting prospects for studies of non-linear effects in the dynamically developing field of cavity-polariton lattices.\cite{StJean:NaturePhot2017,Su:NatPhys2020} In particular, the double microcavity structure should enable observation of three-dimensional polariton hopping, with a transfer occurring in two directions: parallel and perpendicular to the microcavities' plane. Apart from the novelty value for fundamental research, the present results are also of high importance for practical applications, e.g., a realization of tunable multi-wavelength sources of coherent light with an ultra-low lasing threshold or development of integrated, all-optical devices performing logic operations.\cite{Yamamoto:NPJQInf2017,Opala:PRB2018, Ballarini:NanoLett2020} For instance, excitation power dependent intensity switching and blueshift of signals from the two lowest polariton levels provide two outputs, enabling the realization of all-optical two-qubit gates~\cite{Xue:PRR2021}.

\section{Conclusions}

We have observed non-resonantly driven polariton condensation and lasing in a four-level system formed in a strongly coupled double microcavity structure. Emission from the two lowest polariton branches features non-linear behaviour with a common threshold in input-output characteristics. The time-resolved measurements reveal that condensates at these two branches do not form simultaneously following the excitation pulse, but rather emerge and decay subsequently one after the other. We have introduced a rate-equation model that properly describes the time-resolved and time-integrated emission intensities and gives insight into the processes governing the polariton dynamics in a multilevel system. Moreover, we show that with the formation of the condensate in the upper of the two lowest branches, energy-degenerate polariton parametric scattering from the upper to the lower branch is launched. This indicates that the condensate created by a non-resonant excitation may replace the resonant pump used previously.

%%%%%%%
\bibliography{KSawicki_MCs_Manuscript_Polariton_lasing}

\begin{thebibliography}{61}
\expandafter\ifx\csname natexlab\endcsname\relax\def\natexlab#1{#1}\fi
\expandafter\ifx\csname bibnamefont\endcsname\relax
  \def\bibnamefont#1{#1}\fi
\expandafter\ifx\csname bibfnamefont\endcsname\relax
  \def\bibfnamefont#1{#1}\fi
\expandafter\ifx\csname citenamefont\endcsname\relax
  \def\citenamefont#1{#1}\fi
\expandafter\ifx\csname url\endcsname\relax
  \def\url#1{\texttt{#1}}\fi
\expandafter\ifx\csname urlprefix\endcsname\relax\def\urlprefix{URL }\fi
\providecommand{\bibinfo}[2]{#2}
\providecommand{\eprint}[2][]{\url{#2}}

\bibitem[{\citenamefont{Weisbuch et~al.}(1992)\citenamefont{Weisbuch, Nishioka,
  Ishikawa, and Arakawa}}]{Weisbuch:PRL1992}
\bibinfo{author}{\bibfnamefont{C.}~\bibnamefont{Weisbuch}},
  \bibinfo{author}{\bibfnamefont{M.}~\bibnamefont{Nishioka}},
  \bibinfo{author}{\bibfnamefont{A.}~\bibnamefont{Ishikawa}}, \bibnamefont{and}
  \bibinfo{author}{\bibfnamefont{Y.}~\bibnamefont{Arakawa}},
  \emph{\bibinfo{title}{Observation of the coupled exciton-photon mode
  splitting in a semiconductor quantum microcavity}}, \bibinfo{journal}{Phys.
  Rev. Lett.} \textbf{\bibinfo{volume}{69}}, \bibinfo{pages}{3314 }
  (\bibinfo{year}{1992}).

\bibitem[{\citenamefont{Ciuti et~al.}(2000)\citenamefont{Ciuti, Schwendimann,
  Deveaud, and Quattropani}}]{Ciuti:PRB2000}
\bibinfo{author}{\bibfnamefont{C.}~\bibnamefont{Ciuti}},
  \bibinfo{author}{\bibfnamefont{P.}~\bibnamefont{Schwendimann}},
  \bibinfo{author}{\bibfnamefont{B.}~\bibnamefont{Deveaud}}, \bibnamefont{and}
  \bibinfo{author}{\bibfnamefont{A.}~\bibnamefont{Quattropani}},
  \emph{\bibinfo{title}{Theory of the angle-resonant polariton amplifier}},
  \bibinfo{journal}{Phys. Rev. B} \textbf{\bibinfo{volume}{62}},
  \bibinfo{pages}{R4825} (\bibinfo{year}{2000}).

\bibitem[{\citenamefont{Krizhanovskii et~al.}(2007)\citenamefont{Krizhanovskii,
  Love, Sanvitto, Whittaker, Skolnick, and Roberts}}]{Krizhanovskii:PRB2007}
\bibinfo{author}{\bibfnamefont{D.~N.} \bibnamefont{Krizhanovskii}},
  \bibinfo{author}{\bibfnamefont{A.~P.~D.} \bibnamefont{Love}},
  \bibinfo{author}{\bibfnamefont{D.}~\bibnamefont{Sanvitto}},
  \bibinfo{author}{\bibfnamefont{D.~M.} \bibnamefont{Whittaker}},
  \bibinfo{author}{\bibfnamefont{M.~S.} \bibnamefont{Skolnick}},
  \bibnamefont{and} \bibinfo{author}{\bibfnamefont{J.~S.}
  \bibnamefont{Roberts}}, \emph{\bibinfo{title}{Interaction between a
  high-density polariton phase and the exciton environment in semiconductor
  microcavities}}, \bibinfo{journal}{Phys. Rev. B}
  \textbf{\bibinfo{volume}{75}}, \bibinfo{pages}{233307}
  (\bibinfo{year}{2007}).

\bibitem[{\citenamefont{Amo et~al.}(2009)\citenamefont{Amo, Lefr\`{e}re,
  Pigeon, Adrados, Ciuti, Carusotto, Houdr\'e, Giacobino, and
  Bramati}}]{Amo:NatPhys2009}
\bibinfo{author}{\bibfnamefont{A.}~\bibnamefont{Amo}},
  \bibinfo{author}{\bibfnamefont{J.}~\bibnamefont{Lefr\`{e}re}},
  \bibinfo{author}{\bibfnamefont{S.}~\bibnamefont{Pigeon}},
  \bibinfo{author}{\bibfnamefont{C.}~\bibnamefont{Adrados}},
  \bibinfo{author}{\bibfnamefont{C.}~\bibnamefont{Ciuti}},
  \bibinfo{author}{\bibfnamefont{I.}~\bibnamefont{Carusotto}},
  \bibinfo{author}{\bibfnamefont{R.}~\bibnamefont{Houdr\'e}},
  \bibinfo{author}{\bibfnamefont{E.}~\bibnamefont{Giacobino}},
  \bibnamefont{and} \bibinfo{author}{\bibfnamefont{A.}~\bibnamefont{Bramati}},
  \emph{\bibinfo{title}{Superfluidity of polaritons in semiconductor
  microcavities}}, \bibinfo{journal}{Nat. Phys.} \textbf{\bibinfo{volume}{5}},
  \bibinfo{pages}{805 } (\bibinfo{year}{2009}).

\bibitem[{\citenamefont{Savvidis et~al.}(2000)\citenamefont{Savvidis, Baumberg,
  Stevenson, Skolnick, Whittaker, and Roberts}}]{Savvidis:PRL2000}
\bibinfo{author}{\bibfnamefont{P.~G.} \bibnamefont{Savvidis}},
  \bibinfo{author}{\bibfnamefont{J.~J.} \bibnamefont{Baumberg}},
  \bibinfo{author}{\bibfnamefont{R.~M.} \bibnamefont{Stevenson}},
  \bibinfo{author}{\bibfnamefont{M.~S.} \bibnamefont{Skolnick}},
  \bibinfo{author}{\bibfnamefont{D.~M.} \bibnamefont{Whittaker}},
  \bibnamefont{and} \bibinfo{author}{\bibfnamefont{J.~S.}
  \bibnamefont{Roberts}}, \emph{\bibinfo{title}{Angle-Resonant Stimulated
  Polariton Amplifier}}, \bibinfo{journal}{Phys. Rev. Lett.}
  \textbf{\bibinfo{volume}{84}}, \bibinfo{pages}{1547} (\bibinfo{year}{2000}).

\bibitem[{\citenamefont{Deng et~al.}(2002)\citenamefont{Deng, Weihs, Santori,
  Bloch, and Yamamoto}}]{Deng:Science2002}
\bibinfo{author}{\bibfnamefont{H.}~\bibnamefont{Deng}},
  \bibinfo{author}{\bibfnamefont{G.}~\bibnamefont{Weihs}},
  \bibinfo{author}{\bibfnamefont{C.}~\bibnamefont{Santori}},
  \bibinfo{author}{\bibfnamefont{J.}~\bibnamefont{Bloch}}, \bibnamefont{and}
  \bibinfo{author}{\bibfnamefont{Y.}~\bibnamefont{Yamamoto}},
  \emph{\bibinfo{title}{Condensation of Semiconductor Microcavity Exciton
  Polaritons}}, \bibinfo{journal}{Science} \textbf{\bibinfo{volume}{298}},
  \bibinfo{pages}{199 } (\bibinfo{year}{2002}).

\bibitem[{\citenamefont{Kasprzak et~al.}(2006)\citenamefont{Kasprzak, Richard,
  Kundermann, Baas, Jeambrun, Keeling, Marchetti, Szyma{\'n}ska, Andr{\'e},
  Staehli et~al.}}]{Kasprzak:Nature2006}
\bibinfo{author}{\bibfnamefont{J.}~\bibnamefont{Kasprzak}},
  \bibinfo{author}{\bibfnamefont{M.}~\bibnamefont{Richard}},
  \bibinfo{author}{\bibfnamefont{S.}~\bibnamefont{Kundermann}},
  \bibinfo{author}{\bibfnamefont{A.}~\bibnamefont{Baas}},
  \bibinfo{author}{\bibfnamefont{P.}~\bibnamefont{Jeambrun}},
  \bibinfo{author}{\bibfnamefont{J.~M.~J.} \bibnamefont{Keeling}},
  \bibinfo{author}{\bibfnamefont{F.~M.} \bibnamefont{Marchetti}},
  \bibinfo{author}{\bibfnamefont{M.~H.} \bibnamefont{Szyma{\'n}ska}},
  \bibinfo{author}{\bibfnamefont{R.}~\bibnamefont{Andr{\'e}}},
  \bibinfo{author}{\bibfnamefont{J.~L.} \bibnamefont{Staehli}},
  \bibinfo{author}{\bibfnamefont{V.}~\bibnamefont{Savona}},
  \bibinfo{author}{\bibfnamefont{P.~B.} \bibnamefont{Littlewood}},
  \bibinfo{author}{\bibfnamefont{D.}~\bibnamefont{B.}}, \bibnamefont{and}
  \bibinfo{author}{\bibfnamefont{L.~S.} \bibnamefont{Dang}},
  \emph{\bibinfo{title}{Bose-Einstein condensation of exciton polaritons}},
  \bibinfo{journal}{Nature} \textbf{\bibinfo{volume}{443}}, \bibinfo{pages}{409
  } (\bibinfo{year}{2006}).

\bibitem[{\citenamefont{Anderson et~al.}(1995)\citenamefont{Anderson, Ensher,
  Matthews, Wieman, and A.}}]{Anderson:Science1995}
\bibinfo{author}{\bibfnamefont{M.~H.} \bibnamefont{Anderson}},
  \bibinfo{author}{\bibfnamefont{J.~R.} \bibnamefont{Ensher}},
  \bibinfo{author}{\bibfnamefont{M.~R.} \bibnamefont{Matthews}},
  \bibinfo{author}{\bibfnamefont{C.~E.} \bibnamefont{Wieman}},
  \bibnamefont{and} \bibinfo{author}{\bibfnamefont{C.~E.} \bibnamefont{A.}},
  \emph{\bibinfo{title}{Observation of Bose-Einstein Condensation in a Dilute
  Atomic Vapor}}, \bibinfo{journal}{Science} \textbf{\bibinfo{volume}{269}},
  \bibinfo{pages}{198} (\bibinfo{year}{1995}).

\bibitem[{\citenamefont{Imamoglu et~al.}(1996)\citenamefont{Imamoglu, Ram, Pau,
  and Yamamoto}}]{Imamoglu:PRA1996}
\bibinfo{author}{\bibfnamefont{A.}~\bibnamefont{Imamoglu}},
  \bibinfo{author}{\bibfnamefont{R.~J.} \bibnamefont{Ram}},
  \bibinfo{author}{\bibfnamefont{S.}~\bibnamefont{Pau}}, \bibnamefont{and}
  \bibinfo{author}{\bibfnamefont{Y.}~\bibnamefont{Yamamoto}},
  \emph{\bibinfo{title}{Nonequilibrium condensates and lasers without
  inversion: Exciton-polariton lasers}}, \bibinfo{journal}{Phys. Rev. A}
  \textbf{\bibinfo{volume}{53}}, \bibinfo{pages}{4250 } (\bibinfo{year}{1996}).

\bibitem[{\citenamefont{Deng et~al.}(2003)\citenamefont{Deng, Weihs, Snoke,
  Bloch, and Yamamoto}}]{Deng:PNAS2003}
\bibinfo{author}{\bibfnamefont{H.}~\bibnamefont{Deng}},
  \bibinfo{author}{\bibfnamefont{G.}~\bibnamefont{Weihs}},
  \bibinfo{author}{\bibfnamefont{D.}~\bibnamefont{Snoke}},
  \bibinfo{author}{\bibfnamefont{J.}~\bibnamefont{Bloch}}, \bibnamefont{and}
  \bibinfo{author}{\bibfnamefont{Y.}~\bibnamefont{Yamamoto}},
  \emph{\bibinfo{title}{Polariton lasing vs. photon lasing in a semiconductor
  microcavity}}, \bibinfo{journal}{Proc. Natl. Acad. Sci. U. S. A.}
  \textbf{\bibinfo{volume}{100}}, \bibinfo{pages}{15318 }
  (\bibinfo{year}{2003}).

\bibitem[{\citenamefont{Christopoulos et~al.}(2007)\citenamefont{Christopoulos,
  von H\"ogersthal, Grundy, Lagoudakis, Kavokin, Baumberg, Christmann, Butt\'e,
  Feltin, Carlin et~al.}}]{Christopoulos:PRL2007}
\bibinfo{author}{\bibfnamefont{S.}~\bibnamefont{Christopoulos}},
  \bibinfo{author}{\bibfnamefont{G.~B.~H.} \bibnamefont{von H\"ogersthal}},
  \bibinfo{author}{\bibfnamefont{A.~J.~D.} \bibnamefont{Grundy}},
  \bibinfo{author}{\bibfnamefont{P.~G.} \bibnamefont{Lagoudakis}},
  \bibinfo{author}{\bibfnamefont{A.~V.} \bibnamefont{Kavokin}},
  \bibinfo{author}{\bibfnamefont{J.~J.} \bibnamefont{Baumberg}},
  \bibinfo{author}{\bibfnamefont{G.}~\bibnamefont{Christmann}},
  \bibinfo{author}{\bibfnamefont{R.}~\bibnamefont{Butt\'e}},
  \bibinfo{author}{\bibfnamefont{E.}~\bibnamefont{Feltin}},
  \bibinfo{author}{\bibfnamefont{J.-F.} \bibnamefont{Carlin}},
  \bibnamefont{and}
  \bibinfo{author}{\bibfnamefont{N.}~\bibnamefont{Grandjean}},
  \emph{\bibinfo{title}{Room-Temperature Polariton Lasing in Semiconductor
  Microcavities}}, \bibinfo{journal}{Phys. Rev. Lett.}
  \textbf{\bibinfo{volume}{98}}, \bibinfo{pages}{126405}
  (\bibinfo{year}{2007}).

\bibitem[{\citenamefont{Bajoni et~al.}(2008)\citenamefont{Bajoni, Senellart,
  Wertz, Sagnes, Miard, Lema\^{\i}tre, and Bloch}}]{Bajoni:PRL2008}
\bibinfo{author}{\bibfnamefont{D.}~\bibnamefont{Bajoni}},
  \bibinfo{author}{\bibfnamefont{P.}~\bibnamefont{Senellart}},
  \bibinfo{author}{\bibfnamefont{E.}~\bibnamefont{Wertz}},
  \bibinfo{author}{\bibfnamefont{I.}~\bibnamefont{Sagnes}},
  \bibinfo{author}{\bibfnamefont{A.}~\bibnamefont{Miard}},
  \bibinfo{author}{\bibfnamefont{A.}~\bibnamefont{Lema\^{\i}tre}},
  \bibnamefont{and} \bibinfo{author}{\bibfnamefont{J.}~\bibnamefont{Bloch}},
  \emph{\bibinfo{title}{Polariton Laser Using Single Micropillar
  $\mathrm{GaAs}\mathrm{\text{\ensuremath{-}}}\mathrm{GaAlAs}$ Semiconductor
  Cavities}}, \bibinfo{journal}{Phys. Rev. Lett.}
  \textbf{\bibinfo{volume}{100}}, \bibinfo{pages}{047401}
  (\bibinfo{year}{2008}).

\bibitem[{\citenamefont{Galbiati et~al.}(2012)\citenamefont{Galbiati, Ferrier,
  Solnyshkov, Tanese, Wertz, Amo, Abbarchi, Senellart, Sagnes, Lema\^{\i}tre
  et~al.}}]{Galbiati:PRL2012}
\bibinfo{author}{\bibfnamefont{M.}~\bibnamefont{Galbiati}},
  \bibinfo{author}{\bibfnamefont{L.}~\bibnamefont{Ferrier}},
  \bibinfo{author}{\bibfnamefont{D.~D.} \bibnamefont{Solnyshkov}},
  \bibinfo{author}{\bibfnamefont{D.}~\bibnamefont{Tanese}},
  \bibinfo{author}{\bibfnamefont{E.}~\bibnamefont{Wertz}},
  \bibinfo{author}{\bibfnamefont{A.}~\bibnamefont{Amo}},
  \bibinfo{author}{\bibfnamefont{M.}~\bibnamefont{Abbarchi}},
  \bibinfo{author}{\bibfnamefont{P.}~\bibnamefont{Senellart}},
  \bibinfo{author}{\bibfnamefont{I.}~\bibnamefont{Sagnes}},
  \bibinfo{author}{\bibfnamefont{A.}~\bibnamefont{Lema\^{\i}tre}},
  \bibinfo{author}{\bibfnamefont{E.}~\bibnamefont{Galopin}},
  \bibinfo{author}{\bibfnamefont{G.}~\bibnamefont{Malpuech}}, \bibnamefont{and}
  \bibinfo{author}{\bibfnamefont{J.}~\bibnamefont{Bloch}},
  \emph{\bibinfo{title}{Polariton Condensation in Photonic Molecules}},
  \bibinfo{journal}{Phys. Rev. Lett.} \textbf{\bibinfo{volume}{108}},
  \bibinfo{pages}{126403} (\bibinfo{year}{2012}).

\bibitem[{\citenamefont{Grosso et~al.}(2014)\citenamefont{Grosso, Trebaol,
  Wouters, Morier-Genoud, Portella-Oberli, and Deveaud}}]{Grosso:PRB2014}
\bibinfo{author}{\bibfnamefont{G.}~\bibnamefont{Grosso}},
  \bibinfo{author}{\bibfnamefont{S.}~\bibnamefont{Trebaol}},
  \bibinfo{author}{\bibfnamefont{M.}~\bibnamefont{Wouters}},
  \bibinfo{author}{\bibfnamefont{F.}~\bibnamefont{Morier-Genoud}},
  \bibinfo{author}{\bibfnamefont{M.~T.} \bibnamefont{Portella-Oberli}},
  \bibnamefont{and} \bibinfo{author}{\bibfnamefont{B.}~\bibnamefont{Deveaud}},
  \emph{\bibinfo{title}{Nonlinear relaxation and selective polychromatic lasing
  of confined polaritons}}, \bibinfo{journal}{Phys. Rev. B}
  \textbf{\bibinfo{volume}{90}}, \bibinfo{pages}{045307}
  (\bibinfo{year}{2014}).

\bibitem[{\citenamefont{Klein et~al.}(2015)\citenamefont{Klein, Klembt, Durupt,
  Kruse, Hommel, and Richard}}]{Klein:APL2015}
\bibinfo{author}{\bibfnamefont{T.}~\bibnamefont{Klein}},
  \bibinfo{author}{\bibfnamefont{S.}~\bibnamefont{Klembt}},
  \bibinfo{author}{\bibfnamefont{E.}~\bibnamefont{Durupt}},
  \bibinfo{author}{\bibfnamefont{C.}~\bibnamefont{Kruse}},
  \bibinfo{author}{\bibfnamefont{D.}~\bibnamefont{Hommel}}, \bibnamefont{and}
  \bibinfo{author}{\bibfnamefont{M.}~\bibnamefont{Richard}},
  \emph{\bibinfo{title}{Polariton lasing in high-quality selenide-based
  micropillars in the strong coupling regime}}, \bibinfo{journal}{Appl. Phys.
  Lett.} \textbf{\bibinfo{volume}{107}}, \bibinfo{pages}{071101}
  (\bibinfo{year}{2015}).

\bibitem[{\citenamefont{Rodriguez et~al.}(2016)\citenamefont{Rodriguez, Amo,
  Sagnes, Le~Gratiet, Galopin, Lema{\^i}tre, and
  Bloch}}]{Rodriguez:NatComm2016}
\bibinfo{author}{\bibfnamefont{S.~R.~K.} \bibnamefont{Rodriguez}},
  \bibinfo{author}{\bibfnamefont{A.}~\bibnamefont{Amo}},
  \bibinfo{author}{\bibfnamefont{I.}~\bibnamefont{Sagnes}},
  \bibinfo{author}{\bibfnamefont{L.}~\bibnamefont{Le~Gratiet}},
  \bibinfo{author}{\bibfnamefont{E.}~\bibnamefont{Galopin}},
  \bibinfo{author}{\bibfnamefont{A.}~\bibnamefont{Lema{\^i}tre}},
  \bibnamefont{and} \bibinfo{author}{\bibfnamefont{J.}~\bibnamefont{Bloch}},
  \emph{\bibinfo{title}{Interaction-induced hopping phase in driven-dissipative
  coupled photonic microcavities}}, \bibinfo{journal}{Nat. Commun.}
  \textbf{\bibinfo{volume}{7}}, \bibinfo{pages}{1} (\bibinfo{year}{2016}).

\bibitem[{\citenamefont{Su et~al.}(2017)\citenamefont{Su, Diederichs, Wang,
  Liew, Zhao, Liu, Xu, Chen, and Xiong}}]{Su:NanoLett2017}
\bibinfo{author}{\bibfnamefont{R.}~\bibnamefont{Su}},
  \bibinfo{author}{\bibfnamefont{C.}~\bibnamefont{Diederichs}},
  \bibinfo{author}{\bibfnamefont{J.}~\bibnamefont{Wang}},
  \bibinfo{author}{\bibfnamefont{T.~C.~H.} \bibnamefont{Liew}},
  \bibinfo{author}{\bibfnamefont{J.}~\bibnamefont{Zhao}},
  \bibinfo{author}{\bibfnamefont{S.}~\bibnamefont{Liu}},
  \bibinfo{author}{\bibfnamefont{W.}~\bibnamefont{Xu}},
  \bibinfo{author}{\bibfnamefont{Z.}~\bibnamefont{Chen}}, \bibnamefont{and}
  \bibinfo{author}{\bibfnamefont{Q.}~\bibnamefont{Xiong}},
  \emph{\bibinfo{title}{Room-Temperature Polariton Lasing in All-Inorganic
  Perovskite Nanoplatelets}}, \bibinfo{journal}{Nano Lett.}
  \textbf{\bibinfo{volume}{17}}, \bibinfo{pages}{3982} (\bibinfo{year}{2017}).

\bibitem[{\citenamefont{Su et~al.}(2020)\citenamefont{Su, Ghosh, Wang, Liu,
  Diederichs, Liew, and Xiong}}]{Su:NatPhys2020}
\bibinfo{author}{\bibfnamefont{R.}~\bibnamefont{Su}},
  \bibinfo{author}{\bibfnamefont{S.}~\bibnamefont{Ghosh}},
  \bibinfo{author}{\bibfnamefont{J.}~\bibnamefont{Wang}},
  \bibinfo{author}{\bibfnamefont{S.}~\bibnamefont{Liu}},
  \bibinfo{author}{\bibfnamefont{C.}~\bibnamefont{Diederichs}},
  \bibinfo{author}{\bibfnamefont{T.~C.~H.} \bibnamefont{Liew}},
  \bibnamefont{and} \bibinfo{author}{\bibfnamefont{Q.}~\bibnamefont{Xiong}},
  \emph{\bibinfo{title}{Observation of exciton polariton condensation in a
  perovskite lattice at room temperature}}, \bibinfo{journal}{Nat. Phys.}
  \textbf{\bibinfo{volume}{16}}, \bibinfo{pages}{301 } (\bibinfo{year}{2020}).

\bibitem[{\citenamefont{Carlon~Zambon et~al.}(2020)\citenamefont{Carlon~Zambon,
  Rodriguez, Lema\^{\i}tre, Harouri, Le~Gratiet, Sagnes, St-Jean, Ravets, Amo,
  and Bloch}}]{Zambon:PRA2020}
\bibinfo{author}{\bibfnamefont{N.}~\bibnamefont{Carlon~Zambon}},
  \bibinfo{author}{\bibfnamefont{S.~R.~K.} \bibnamefont{Rodriguez}},
  \bibinfo{author}{\bibfnamefont{A.}~\bibnamefont{Lema\^{\i}tre}},
  \bibinfo{author}{\bibfnamefont{A.}~\bibnamefont{Harouri}},
  \bibinfo{author}{\bibfnamefont{L.}~\bibnamefont{Le~Gratiet}},
  \bibinfo{author}{\bibfnamefont{I.}~\bibnamefont{Sagnes}},
  \bibinfo{author}{\bibfnamefont{P.}~\bibnamefont{St-Jean}},
  \bibinfo{author}{\bibfnamefont{S.}~\bibnamefont{Ravets}},
  \bibinfo{author}{\bibfnamefont{A.}~\bibnamefont{Amo}}, \bibnamefont{and}
  \bibinfo{author}{\bibfnamefont{J.}~\bibnamefont{Bloch}},
  \emph{\bibinfo{title}{Parametric instability in coupled nonlinear
  microcavities}}, \bibinfo{journal}{Phys. Rev. A}
  \textbf{\bibinfo{volume}{102}}, \bibinfo{pages}{023526}
  (\bibinfo{year}{2020}).

\bibitem[{\citenamefont{Ardizzone et~al.}(2013)\citenamefont{Ardizzone,
  Lewandowski, Luk, Tse, Kwong, L{\"u}cke, Abbarchi, Baudin, Galopin, Bloch
  et~al.}}]{Ardizzone:SciRep2013}
\bibinfo{author}{\bibfnamefont{V.}~\bibnamefont{Ardizzone}},
  \bibinfo{author}{\bibfnamefont{P.}~\bibnamefont{Lewandowski}},
  \bibinfo{author}{\bibfnamefont{M.-H.} \bibnamefont{Luk}},
  \bibinfo{author}{\bibfnamefont{Y.-C.} \bibnamefont{Tse}},
  \bibinfo{author}{\bibfnamefont{N.-H.} \bibnamefont{Kwong}},
  \bibinfo{author}{\bibfnamefont{A.}~\bibnamefont{L{\"u}cke}},
  \bibinfo{author}{\bibfnamefont{M.}~\bibnamefont{Abbarchi}},
  \bibinfo{author}{\bibfnamefont{E.}~\bibnamefont{Baudin}},
  \bibinfo{author}{\bibfnamefont{E.}~\bibnamefont{Galopin}},
  \bibinfo{author}{\bibfnamefont{J.}~\bibnamefont{Bloch}},
  \bibinfo{author}{\bibfnamefont{A.}~\bibnamefont{Lemaitre}},
  \bibinfo{author}{\bibfnamefont{P.~T.} \bibnamefont{Leung}},
  \bibinfo{author}{\bibfnamefont{P.}~\bibnamefont{Roussignol}},
  \bibinfo{author}{\bibfnamefont{R.}~\bibnamefont{Binder}},
  \bibinfo{author}{\bibfnamefont{J.}~\bibnamefont{Tignon}}, \bibnamefont{and}
  \bibinfo{author}{\bibfnamefont{S.}~\bibnamefont{Schumacher}},
  \emph{\bibinfo{title}{Formation and control of Turing patterns in a coherent
  quantum fluid}}, \bibinfo{journal}{Scientific Reports}
  \textbf{\bibinfo{volume}{3}}, \bibinfo{pages}{3016} (\bibinfo{year}{2013}).

\bibitem[{\citenamefont{Diederichs et~al.}(2006)\citenamefont{Diederichs,
  Tignon, Dasbach, Ciuti, Lema{\^i}tre, Bloch, Roussignol, and
  Delalande}}]{Diederichs:Nature2006}
\bibinfo{author}{\bibfnamefont{C.}~\bibnamefont{Diederichs}},
  \bibinfo{author}{\bibfnamefont{J.}~\bibnamefont{Tignon}},
  \bibinfo{author}{\bibfnamefont{G.}~\bibnamefont{Dasbach}},
  \bibinfo{author}{\bibfnamefont{C.}~\bibnamefont{Ciuti}},
  \bibinfo{author}{\bibfnamefont{A.}~\bibnamefont{Lema{\^i}tre}},
  \bibinfo{author}{\bibfnamefont{J.}~\bibnamefont{Bloch}},
  \bibinfo{author}{\bibfnamefont{P.}~\bibnamefont{Roussignol}},
  \bibnamefont{and}
  \bibinfo{author}{\bibfnamefont{C.}~\bibnamefont{Delalande}},
  \emph{\bibinfo{title}{Parametric oscillation in vertical triple
  microcavities}}, \bibinfo{journal}{Nature} \textbf{\bibinfo{volume}{440}},
  \bibinfo{pages}{904} (\bibinfo{year}{2006}).

\bibitem[{\citenamefont{Hartmann et~al.}(2006)\citenamefont{Hartmann, Brandao,
  and Plenio}}]{Hartmann:NaturePhys2006}
\bibinfo{author}{\bibfnamefont{M.~J.} \bibnamefont{Hartmann}},
  \bibinfo{author}{\bibfnamefont{F.~G.} \bibnamefont{Brandao}},
  \bibnamefont{and} \bibinfo{author}{\bibfnamefont{M.~B.}
  \bibnamefont{Plenio}}, \emph{\bibinfo{title}{Strongly interacting polaritons
  in coupled arrays of cavities}}, \bibinfo{journal}{Nat. Phys.}
  \textbf{\bibinfo{volume}{2}}, \bibinfo{pages}{849} (\bibinfo{year}{2006}).

\bibitem[{\citenamefont{Armitage et~al.}(1998)\citenamefont{Armitage, Skolnick,
  Astratov, Whittaker, Panzarini, Andreani, Fisher, Roberts, Kavokin,
  Kaliteevski et~al.}}]{Armitage:PRB1998}
\bibinfo{author}{\bibfnamefont{A.}~\bibnamefont{Armitage}},
  \bibinfo{author}{\bibfnamefont{M.~S.} \bibnamefont{Skolnick}},
  \bibinfo{author}{\bibfnamefont{V.~N.} \bibnamefont{Astratov}},
  \bibinfo{author}{\bibfnamefont{D.~M.} \bibnamefont{Whittaker}},
  \bibinfo{author}{\bibfnamefont{G.}~\bibnamefont{Panzarini}},
  \bibinfo{author}{\bibfnamefont{L.~C.} \bibnamefont{Andreani}},
  \bibinfo{author}{\bibfnamefont{T.~A.} \bibnamefont{Fisher}},
  \bibinfo{author}{\bibfnamefont{J.~S.} \bibnamefont{Roberts}},
  \bibinfo{author}{\bibfnamefont{A.~V.} \bibnamefont{Kavokin}},
  \bibinfo{author}{\bibfnamefont{M.~A.} \bibnamefont{Kaliteevski}},
  \bibnamefont{and} \bibinfo{author}{\bibfnamefont{M.~R.}
  \bibnamefont{Vladimirova}}, \emph{\bibinfo{title}{Optically induced splitting
  of bright excitonic states in coupled quantum microcavities}},
  \bibinfo{journal}{Phys. Rev. B} \textbf{\bibinfo{volume}{57}},
  \bibinfo{pages}{14877 } (\bibinfo{year}{1998}).

\bibitem[{\citenamefont{Chang et~al.}(2014)\citenamefont{Chang, Jiang, Hua,
  Yang, Wen, Jiang, Li, Wang, and Xiao}}]{Chang:NaturePhot2014}
\bibinfo{author}{\bibfnamefont{L.}~\bibnamefont{Chang}},
  \bibinfo{author}{\bibfnamefont{X.}~\bibnamefont{Jiang}},
  \bibinfo{author}{\bibfnamefont{S.}~\bibnamefont{Hua}},
  \bibinfo{author}{\bibfnamefont{C.}~\bibnamefont{Yang}},
  \bibinfo{author}{\bibfnamefont{J.}~\bibnamefont{Wen}},
  \bibinfo{author}{\bibfnamefont{L.}~\bibnamefont{Jiang}},
  \bibinfo{author}{\bibfnamefont{G.}~\bibnamefont{Li}},
  \bibinfo{author}{\bibfnamefont{G.}~\bibnamefont{Wang}}, \bibnamefont{and}
  \bibinfo{author}{\bibfnamefont{M.}~\bibnamefont{Xiao}},
  \emph{\bibinfo{title}{Parity--time symmetry and variable optical isolation in
  active--passive-coupled microresonators}}, \bibinfo{journal}{Nat. Photonics}
  \textbf{\bibinfo{volume}{8}}, \bibinfo{pages}{524} (\bibinfo{year}{2014}).

\bibitem[{\citenamefont{Hamel et~al.}(2015)\citenamefont{Hamel, Haddadi,
  Raineri, Monnier, Beaudoin, Sagnes, Levenson, and
  Yacomotti}}]{Hamel:NatPhot2015}
\bibinfo{author}{\bibfnamefont{P.}~\bibnamefont{Hamel}},
  \bibinfo{author}{\bibfnamefont{S.}~\bibnamefont{Haddadi}},
  \bibinfo{author}{\bibfnamefont{F.}~\bibnamefont{Raineri}},
  \bibinfo{author}{\bibfnamefont{P.}~\bibnamefont{Monnier}},
  \bibinfo{author}{\bibfnamefont{G.}~\bibnamefont{Beaudoin}},
  \bibinfo{author}{\bibfnamefont{I.}~\bibnamefont{Sagnes}},
  \bibinfo{author}{\bibfnamefont{A.}~\bibnamefont{Levenson}}, \bibnamefont{and}
  \bibinfo{author}{\bibfnamefont{A.~M.} \bibnamefont{Yacomotti}},
  \emph{\bibinfo{title}{Spontaneous mirror-symmetry breaking in coupled
  photonic-crystal nanolasers}}, \bibinfo{journal}{Nat. Phot.}
  \textbf{\bibinfo{volume}{9}}, \bibinfo{pages}{311 } (\bibinfo{year}{2015}).

\bibitem[{\citenamefont{Gerace and Savona}(2014)}]{Gerace:PRA2014}
\bibinfo{author}{\bibfnamefont{D.}~\bibnamefont{Gerace}} \bibnamefont{and}
  \bibinfo{author}{\bibfnamefont{V.}~\bibnamefont{Savona}},
  \emph{\bibinfo{title}{Unconventional photon blockade in doubly resonant
  microcavities with second-order nonlinearity}}, \bibinfo{journal}{Phys. Rev.
  A} \textbf{\bibinfo{volume}{89}}, \bibinfo{pages}{031803}
  (\bibinfo{year}{2014}).

\bibitem[{\citenamefont{Portolan et~al.}(2014)\citenamefont{Portolan,
  Einkemmer, V\"{o}r\"{o}s, Weihs, and Rabl}}]{Portolan:NJofPhys2014}
\bibinfo{author}{\bibfnamefont{S.}~\bibnamefont{Portolan}},
  \bibinfo{author}{\bibfnamefont{L.}~\bibnamefont{Einkemmer}},
  \bibinfo{author}{\bibfnamefont{Z.}~\bibnamefont{V\"{o}r\"{o}s}},
  \bibinfo{author}{\bibfnamefont{G.}~\bibnamefont{Weihs}}, \bibnamefont{and}
  \bibinfo{author}{\bibfnamefont{P.}~\bibnamefont{Rabl}},
  \emph{\bibinfo{title}{Generation of hyper-entangled photon pairs in coupled
  microcavities}}, \bibinfo{journal}{New J. Phys.}
  \textbf{\bibinfo{volume}{16}}, \bibinfo{pages}{063030}
  (\bibinfo{year}{2014}).

\bibitem[{\citenamefont{Stanley et~al.}(1994)\citenamefont{Stanley, Houdr\'e,
  Oesterle, Ilegems, and Weisbuch}}]{Stanley:APL1994}
\bibinfo{author}{\bibfnamefont{R.~P.} \bibnamefont{Stanley}},
  \bibinfo{author}{\bibfnamefont{R.}~\bibnamefont{Houdr\'e}},
  \bibinfo{author}{\bibfnamefont{U.}~\bibnamefont{Oesterle}},
  \bibinfo{author}{\bibfnamefont{M.}~\bibnamefont{Ilegems}}, \bibnamefont{and}
  \bibinfo{author}{\bibfnamefont{C.}~\bibnamefont{Weisbuch}},
  \emph{\bibinfo{title}{Coupled semiconductor microcavities}},
  \bibinfo{journal}{Appl. Phys. Lett.} \textbf{\bibinfo{volume}{65}},
  \bibinfo{pages}{2093} (\bibinfo{year}{1994}).

\bibitem[{\citenamefont{Xie et~al.}(2012)\citenamefont{Xie, Dong, Zhang, Sun,
  Zhou, Ling, Lu, Shen, and Chen}}]{Xie:PRL2012}
\bibinfo{author}{\bibfnamefont{W.}~\bibnamefont{Xie}},
  \bibinfo{author}{\bibfnamefont{H.}~\bibnamefont{Dong}},
  \bibinfo{author}{\bibfnamefont{S.}~\bibnamefont{Zhang}},
  \bibinfo{author}{\bibfnamefont{L.}~\bibnamefont{Sun}},
  \bibinfo{author}{\bibfnamefont{W.}~\bibnamefont{Zhou}},
  \bibinfo{author}{\bibfnamefont{Y.}~\bibnamefont{Ling}},
  \bibinfo{author}{\bibfnamefont{J.}~\bibnamefont{Lu}},
  \bibinfo{author}{\bibfnamefont{X.}~\bibnamefont{Shen}}, \bibnamefont{and}
  \bibinfo{author}{\bibfnamefont{Z.}~\bibnamefont{Chen}},
  \emph{\bibinfo{title}{Room-Temperature Polariton Parametric Scattering Driven
  by a One-Dimensional Polariton Condensate}}, \bibinfo{journal}{Phys. Rev.
  Lett.} \textbf{\bibinfo{volume}{108}}, \bibinfo{pages}{166401}
  (\bibinfo{year}{2012}).

\bibitem[{\citenamefont{Trichet et~al.}(2012)\citenamefont{Trichet,
  M{\'{e}}dard, Z{\'{u}}{\~{n}}iga-P{\'{e}}rez, Alloing, and
  Richard}}]{Trichet:NJP2012}
\bibinfo{author}{\bibfnamefont{A.}~\bibnamefont{Trichet}},
  \bibinfo{author}{\bibfnamefont{F.}~\bibnamefont{M{\'{e}}dard}},
  \bibinfo{author}{\bibfnamefont{J.}~\bibnamefont{Z{\'{u}}{\~{n}}iga-P{\'{e}}rez}},
  \bibinfo{author}{\bibfnamefont{B.}~\bibnamefont{Alloing}}, \bibnamefont{and}
  \bibinfo{author}{\bibfnamefont{M.}~\bibnamefont{Richard}},
  \emph{\bibinfo{title}{From strong to weak coupling regime in a single GaN
  microwire up to room temperature}}, \bibinfo{journal}{New J. Phys.}
  \textbf{\bibinfo{volume}{14}}, \bibinfo{pages}{073004}
  (\bibinfo{year}{2012}).

\bibitem[{\citenamefont{Xu et~al.}(2014)\citenamefont{Xu, Xie, Liu, Wang,
  Zhang, Wang, Zhang, Sun, Shen, and Chen}}]{Xu:APL2014}
\bibinfo{author}{\bibfnamefont{D.}~\bibnamefont{Xu}},
  \bibinfo{author}{\bibfnamefont{W.}~\bibnamefont{Xie}},
  \bibinfo{author}{\bibfnamefont{W.}~\bibnamefont{Liu}},
  \bibinfo{author}{\bibfnamefont{J.}~\bibnamefont{Wang}},
  \bibinfo{author}{\bibfnamefont{L.}~\bibnamefont{Zhang}},
  \bibinfo{author}{\bibfnamefont{Y.}~\bibnamefont{Wang}},
  \bibinfo{author}{\bibfnamefont{S.}~\bibnamefont{Zhang}},
  \bibinfo{author}{\bibfnamefont{L.}~\bibnamefont{Sun}},
  \bibinfo{author}{\bibfnamefont{X.}~\bibnamefont{Shen}}, \bibnamefont{and}
  \bibinfo{author}{\bibfnamefont{Z.}~\bibnamefont{Chen}},
  \emph{\bibinfo{title}{Polariton lasing in a ZnO microwire above 450 K}},
  \bibinfo{journal}{Appl. Phys. Lett.} \textbf{\bibinfo{volume}{104}},
  \bibinfo{pages}{082101} (\bibinfo{year}{2014}).

\bibitem[{\citenamefont{Zhang et~al.}(2018)\citenamefont{Zhang, Zhang, Tang,
  Tian, Xu, Dong, and Zhou}}]{Zhang:Nanoscale2018}
\bibinfo{author}{\bibfnamefont{Y.}~\bibnamefont{Zhang}},
  \bibinfo{author}{\bibfnamefont{X.}~\bibnamefont{Zhang}},
  \bibinfo{author}{\bibfnamefont{B.}~\bibnamefont{Tang}},
  \bibinfo{author}{\bibfnamefont{C.}~\bibnamefont{Tian}},
  \bibinfo{author}{\bibfnamefont{C.}~\bibnamefont{Xu}},
  \bibinfo{author}{\bibfnamefont{H.}~\bibnamefont{Dong}}, \bibnamefont{and}
  \bibinfo{author}{\bibfnamefont{W.}~\bibnamefont{Zhou}},
  \emph{\bibinfo{title}{Realization of an all-optically controlled dynamic
  superlattice for exciton–polaritons}}, \bibinfo{journal}{Nanoscale}
  \textbf{\bibinfo{volume}{10}}, \bibinfo{pages}{14082 }
  (\bibinfo{year}{2018}).

\bibitem[{\citenamefont{Stelitano et~al.}(2009)\citenamefont{Stelitano,
  De~Luca, Savasta, Monsù~Scolaro, and Patané}}]{Stelitano:APL2009}
\bibinfo{author}{\bibfnamefont{S.}~\bibnamefont{Stelitano}},
  \bibinfo{author}{\bibfnamefont{G.}~\bibnamefont{De~Luca}},
  \bibinfo{author}{\bibfnamefont{S.}~\bibnamefont{Savasta}},
  \bibinfo{author}{\bibfnamefont{L.}~\bibnamefont{Monsù~Scolaro}},
  \bibnamefont{and} \bibinfo{author}{\bibfnamefont{S.}~\bibnamefont{Patané}},
  \emph{\bibinfo{title}{Vertical coupled double organic microcavities}},
  \bibinfo{journal}{Appl. Phys. Lett.} \textbf{\bibinfo{volume}{95}},
  \bibinfo{pages}{093303} (\bibinfo{year}{2009}).

\bibitem[{\citenamefont{Liu et~al.}(2015)\citenamefont{Liu, Rai, Grezmak,
  Twieg, and Singer}}]{Liu:PRB2015}
\bibinfo{author}{\bibfnamefont{B.}~\bibnamefont{Liu}},
  \bibinfo{author}{\bibfnamefont{P.}~\bibnamefont{Rai}},
  \bibinfo{author}{\bibfnamefont{J.}~\bibnamefont{Grezmak}},
  \bibinfo{author}{\bibfnamefont{R.~J.} \bibnamefont{Twieg}}, \bibnamefont{and}
  \bibinfo{author}{\bibfnamefont{K.~D.} \bibnamefont{Singer}},
  \emph{\bibinfo{title}{Coupling of exciton-polaritons in
  $\text{low}\ensuremath{-}Q$ coupled microcavities beyond the rotating wave
  approximation}}, \bibinfo{journal}{Phys. Rev. B}
  \textbf{\bibinfo{volume}{92}}, \bibinfo{pages}{155301}
  (\bibinfo{year}{2015}).

\bibitem[{\citenamefont{\'Sciesiek et~al.}(2017)\citenamefont{\'Sciesiek,
  Pacuski, Rousset, Parli\'nska-Wojtan, Golnik, and
  Suffczy\'nski}}]{Sciesiek:CGD2017}
\bibinfo{author}{\bibfnamefont{M.}~\bibnamefont{\'Sciesiek}},
  \bibinfo{author}{\bibfnamefont{W.}~\bibnamefont{Pacuski}},
  \bibinfo{author}{\bibfnamefont{J.-G.} \bibnamefont{Rousset}},
  \bibinfo{author}{\bibfnamefont{M.}~\bibnamefont{Parli\'nska-Wojtan}},
  \bibinfo{author}{\bibfnamefont{A.}~\bibnamefont{Golnik}}, \bibnamefont{and}
  \bibinfo{author}{\bibfnamefont{J.}~\bibnamefont{Suffczy\'nski}},
  \emph{\bibinfo{title}{Design and control of mode interaction in coupled ZnTe
  optical microcavities}}, \bibinfo{journal}{Cryst. Growth Des.}
  \textbf{\bibinfo{volume}{17}}, \bibinfo{pages}{3716} (\bibinfo{year}{2017}).

\bibitem[{\citenamefont{Jayaprakash et~al.}(2019)\citenamefont{Jayaprakash,
  Georgiou, Coulthard, Askitopoulos, Rajendran, Coles, Musser, Clark, Samuel,
  Turnbull et~al.}}]{Jayaprakash:LSA2019}
\bibinfo{author}{\bibfnamefont{R.}~\bibnamefont{Jayaprakash}},
  \bibinfo{author}{\bibfnamefont{K.}~\bibnamefont{Georgiou}},
  \bibinfo{author}{\bibfnamefont{H.}~\bibnamefont{Coulthard}},
  \bibinfo{author}{\bibfnamefont{A.}~\bibnamefont{Askitopoulos}},
  \bibinfo{author}{\bibfnamefont{S.~K.} \bibnamefont{Rajendran}},
  \bibinfo{author}{\bibfnamefont{D.~M.} \bibnamefont{Coles}},
  \bibinfo{author}{\bibfnamefont{A.~J.} \bibnamefont{Musser}},
  \bibinfo{author}{\bibfnamefont{J.}~\bibnamefont{Clark}},
  \bibinfo{author}{\bibfnamefont{I.~D.} \bibnamefont{Samuel}},
  \bibinfo{author}{\bibfnamefont{G.~A.} \bibnamefont{Turnbull}},
  \bibnamefont{et~al.}, \emph{\bibinfo{title}{A hybrid organic--inorganic
  polariton LED}}, \bibinfo{journal}{Light Sci. Appl.}
  \textbf{\bibinfo{volume}{8}}, \bibinfo{pages}{1} (\bibinfo{year}{2019}).

\bibitem[{\citenamefont{\'Sciesiek et~al.}(2020)\citenamefont{\'Sciesiek,
  Sawicki, Pacuski, Sobczak, Kazimierczuk, Golnik, and
  {Suf\mbox{}fczy\'nski}}}]{Sciesiek:CommunMat2020}
\bibinfo{author}{\bibfnamefont{M.}~\bibnamefont{\'Sciesiek}},
  \bibinfo{author}{\bibfnamefont{K.}~\bibnamefont{Sawicki}},
  \bibinfo{author}{\bibfnamefont{W.}~\bibnamefont{Pacuski}},
  \bibinfo{author}{\bibfnamefont{K.}~\bibnamefont{Sobczak}},
  \bibinfo{author}{\bibfnamefont{T.}~\bibnamefont{Kazimierczuk}},
  \bibinfo{author}{\bibfnamefont{A.}~\bibnamefont{Golnik}}, \bibnamefont{and}
  \bibinfo{author}{\bibfnamefont{J.}~\bibnamefont{{Suf\mbox{}fczy\'nski}}},
  \emph{\bibinfo{title}{Long-distance coupling and energy transfer between
  exciton states in magnetically controlled microcavities}},
  \bibinfo{journal}{Commun. Mater.} \textbf{\bibinfo{volume}{1}},
  \bibinfo{pages}{78} (\bibinfo{year}{2020}).

\bibitem[{\citenamefont{Andr\'{e} and Dang}(1997)}]{Andre:JAP1997}
\bibinfo{author}{\bibfnamefont{R.}~\bibnamefont{Andr\'{e}}} \bibnamefont{and}
  \bibinfo{author}{\bibfnamefont{L.~S.} \bibnamefont{Dang}},
  \emph{\bibinfo{title}{Low-temperature refractive indices of
  Cd$_{1-x}$Mn$_x$Te and Cd$_{1-y}$Mg$_y$Te}}, \bibinfo{journal}{J. Appl.
  Phys.} \textbf{\bibinfo{volume}{82}}, \bibinfo{pages}{5086}
  (\bibinfo{year}{1997}).

\bibitem[{\citenamefont{Rousset et~al.}(2013)\citenamefont{Rousset, Kobak,
  Slupinski, Jakubczyk, Stawicki, Janik, Tokarczyk, Kowalski, Nawrocki, and
  Pacuski}}]{Rousset:JCG2013}
\bibinfo{author}{\bibfnamefont{J.-G.} \bibnamefont{Rousset}},
  \bibinfo{author}{\bibfnamefont{J.}~\bibnamefont{Kobak}},
  \bibinfo{author}{\bibfnamefont{T.}~\bibnamefont{Slupinski}},
  \bibinfo{author}{\bibfnamefont{T.}~\bibnamefont{Jakubczyk}},
  \bibinfo{author}{\bibfnamefont{P.}~\bibnamefont{Stawicki}},
  \bibinfo{author}{\bibfnamefont{E.}~\bibnamefont{Janik}},
  \bibinfo{author}{\bibfnamefont{M.}~\bibnamefont{Tokarczyk}},
  \bibinfo{author}{\bibfnamefont{G.}~\bibnamefont{Kowalski}},
  \bibinfo{author}{\bibfnamefont{M.}~\bibnamefont{Nawrocki}}, \bibnamefont{and}
  \bibinfo{author}{\bibfnamefont{W.}~\bibnamefont{Pacuski}},
  \emph{\bibinfo{title}{MBE growth and characterization of a II--VI distributed
  Bragg reflector and microcavity lattice-matched to MgTe}},
  \bibinfo{journal}{J. Cryst. Growth} \textbf{\bibinfo{volume}{378}},
  \bibinfo{pages}{266} (\bibinfo{year}{2013}).

\bibitem[{\citenamefont{Pacuski et~al.}(2017)\citenamefont{Pacuski, Rousset,
  Delmonte, Jakubczyk, Sobczak, Borysiuk, Sawicki, Janik, and
  Kasprzak}}]{Pacuski:CGD2017}
\bibinfo{author}{\bibfnamefont{W.}~\bibnamefont{Pacuski}},
  \bibinfo{author}{\bibfnamefont{J.-G.} \bibnamefont{Rousset}},
  \bibinfo{author}{\bibfnamefont{V.}~\bibnamefont{Delmonte}},
  \bibinfo{author}{\bibfnamefont{T.}~\bibnamefont{Jakubczyk}},
  \bibinfo{author}{\bibfnamefont{K.}~\bibnamefont{Sobczak}},
  \bibinfo{author}{\bibfnamefont{J.}~\bibnamefont{Borysiuk}},
  \bibinfo{author}{\bibfnamefont{K.}~\bibnamefont{Sawicki}},
  \bibinfo{author}{\bibfnamefont{E.}~\bibnamefont{Janik}}, \bibnamefont{and}
  \bibinfo{author}{\bibfnamefont{J.}~\bibnamefont{Kasprzak}},
  \emph{\bibinfo{title}{Antireflective Photonic Structure for Coherent
  Nonlinear Spectroscopy of Single Magnetic Quantum Dots}},
  \bibinfo{journal}{Cryst. Growth Des.} \textbf{\bibinfo{volume}{17}},
  \bibinfo{pages}{2987 } (\bibinfo{year}{2017}).

\bibitem[{\citenamefont{Richter et~al.}(2015)\citenamefont{Richter, Michalsky,
  Fricke, Sturm, Franke, Grundmann, and Schmidt-Grund}}]{Richter:APL2015}
\bibinfo{author}{\bibfnamefont{S.}~\bibnamefont{Richter}},
  \bibinfo{author}{\bibfnamefont{T.}~\bibnamefont{Michalsky}},
  \bibinfo{author}{\bibfnamefont{L.}~\bibnamefont{Fricke}},
  \bibinfo{author}{\bibfnamefont{C.}~\bibnamefont{Sturm}},
  \bibinfo{author}{\bibfnamefont{H.}~\bibnamefont{Franke}},
  \bibinfo{author}{\bibfnamefont{M.}~\bibnamefont{Grundmann}},
  \bibnamefont{and}
  \bibinfo{author}{\bibfnamefont{R.}~\bibnamefont{Schmidt-Grund}},
  \emph{\bibinfo{title}{Maxwell consideration of polaritonic quasi-particle
  Hamiltonians in multi-level systems}}, \bibinfo{journal}{Appl. Phys. Lett.}
  \textbf{\bibinfo{volume}{107}}, \bibinfo{pages}{231104}
  (\bibinfo{year}{2015}).

\bibitem[{\citenamefont{Sawicki et~al.}(2019)\citenamefont{Sawicki, Rousset,
  Rudniewski, Pacuski, {\'S}ciesiek, Kazimierczuk, Sobczak, Borysiuk, Nawrocki,
  and Suf\mbox{}fczy{\'n}ski}}]{Sawicki:CommPhys2019}
\bibinfo{author}{\bibfnamefont{K.}~\bibnamefont{Sawicki}},
  \bibinfo{author}{\bibfnamefont{J.-G.} \bibnamefont{Rousset}},
  \bibinfo{author}{\bibfnamefont{R.}~\bibnamefont{Rudniewski}},
  \bibinfo{author}{\bibfnamefont{W.}~\bibnamefont{Pacuski}},
  \bibinfo{author}{\bibfnamefont{M.}~\bibnamefont{{\'S}ciesiek}},
  \bibinfo{author}{\bibfnamefont{T.}~\bibnamefont{Kazimierczuk}},
  \bibinfo{author}{\bibfnamefont{K.}~\bibnamefont{Sobczak}},
  \bibinfo{author}{\bibfnamefont{J.}~\bibnamefont{Borysiuk}},
  \bibinfo{author}{\bibfnamefont{M.}~\bibnamefont{Nawrocki}}, \bibnamefont{and}
  \bibinfo{author}{\bibfnamefont{J.}~\bibnamefont{Suf\mbox{}fczy{\'n}ski}},
  \emph{\bibinfo{title}{Triple threshold lasing from a photonic trap in a
  Te/Se-based optical microcavity}}, \bibinfo{journal}{Commun. Phys.}
  \textbf{\bibinfo{volume}{2}}, \bibinfo{pages}{38} (\bibinfo{year}{2019}).

\bibitem[{\citenamefont{Liao et~al.}(2019)\citenamefont{Liao, Ling, Luo, Zhang,
  Wang, and Chen}}]{Liao:APE2019}
\bibinfo{author}{\bibfnamefont{L.}~\bibnamefont{Liao}},
  \bibinfo{author}{\bibfnamefont{Y.}~\bibnamefont{Ling}},
  \bibinfo{author}{\bibfnamefont{S.}~\bibnamefont{Luo}},
  \bibinfo{author}{\bibfnamefont{Z.}~\bibnamefont{Zhang}},
  \bibinfo{author}{\bibfnamefont{J.}~\bibnamefont{Wang}}, \bibnamefont{and}
  \bibinfo{author}{\bibfnamefont{Z.}~\bibnamefont{Chen}},
  \emph{\bibinfo{title}{Propagation of a polariton condensate in a
  one-dimensional microwire at room temperature}}, \bibinfo{journal}{Appl.
  Phys. Express} \textbf{\bibinfo{volume}{12}}, \bibinfo{pages}{052009}
  (\bibinfo{year}{2019}).

\bibitem[{\citenamefont{Wouters et~al.}(2008)\citenamefont{Wouters, Carusotto,
  and Ciuti}}]{Wouters:PRB2008}
\bibinfo{author}{\bibfnamefont{M.}~\bibnamefont{Wouters}},
  \bibinfo{author}{\bibfnamefont{I.}~\bibnamefont{Carusotto}},
  \bibnamefont{and} \bibinfo{author}{\bibfnamefont{C.}~\bibnamefont{Ciuti}},
  \emph{\bibinfo{title}{Spatial and spectral shape of inhomogeneous
  nonequilibrium exciton-polariton condensates}}, \bibinfo{journal}{Phys. Rev.
  B} \textbf{\bibinfo{volume}{77}}, \bibinfo{pages}{115340}
  (\bibinfo{year}{2008}).

\bibitem[{\citenamefont{Lagoudakis et~al.}(2011)\citenamefont{Lagoudakis,
  Manni, Pietka, Wouters, Liew, Savona, Kavokin, Andr\'e, and
  Deveaud-Pl\'edran}}]{Lagoudakis:PRL2011}
\bibinfo{author}{\bibfnamefont{K.~G.} \bibnamefont{Lagoudakis}},
  \bibinfo{author}{\bibfnamefont{F.}~\bibnamefont{Manni}},
  \bibinfo{author}{\bibfnamefont{B.}~\bibnamefont{Pietka}},
  \bibinfo{author}{\bibfnamefont{M.}~\bibnamefont{Wouters}},
  \bibinfo{author}{\bibfnamefont{T.~C.~H.} \bibnamefont{Liew}},
  \bibinfo{author}{\bibfnamefont{V.}~\bibnamefont{Savona}},
  \bibinfo{author}{\bibfnamefont{A.~V.} \bibnamefont{Kavokin}},
  \bibinfo{author}{\bibfnamefont{R.}~\bibnamefont{Andr\'e}}, \bibnamefont{and}
  \bibinfo{author}{\bibfnamefont{B.}~\bibnamefont{Deveaud-Pl\'edran}},
  \emph{\bibinfo{title}{Probing the Dynamics of Spontaneous Quantum Vortices in
  Polariton Superfluids}}, \bibinfo{journal}{Phys. Rev. Lett.}
  \textbf{\bibinfo{volume}{106}}, \bibinfo{pages}{115301}
  (\bibinfo{year}{2011}).

\bibitem[{\citenamefont{Winkler et~al.}(2016)\citenamefont{Winkler, Egorov,
  Savenko, Ma, Estrecho, Gao, Muller, Kamp, Liew, Ostrovskaya
  et~al.}}]{Winkler:PRB2016}
\bibinfo{author}{\bibfnamefont{K.}~\bibnamefont{Winkler}},
  \bibinfo{author}{\bibfnamefont{O.}~\bibnamefont{Egorov}},
  \bibinfo{author}{\bibfnamefont{I.~G.} \bibnamefont{Savenko}},
  \bibinfo{author}{\bibfnamefont{X.}~\bibnamefont{Ma}},
  \bibinfo{author}{\bibfnamefont{E.}~\bibnamefont{Estrecho}},
  \bibinfo{author}{\bibfnamefont{T.}~\bibnamefont{Gao}},
  \bibinfo{author}{\bibfnamefont{S.}~\bibnamefont{Muller}},
  \bibinfo{author}{\bibfnamefont{M.}~\bibnamefont{Kamp}},
  \bibinfo{author}{\bibfnamefont{T.~C.~H.} \bibnamefont{Liew}},
  \bibinfo{author}{\bibfnamefont{E.}~\bibnamefont{Ostrovskaya}},
  \bibinfo{author}{\bibfnamefont{S.}~\bibnamefont{Hofling}}, \bibnamefont{and}
  \bibinfo{author}{\bibfnamefont{C.}~\bibnamefont{Schneider}},
  \emph{\bibinfo{title}{Collective state transitions of exciton-polaritons
  loaded into a periodic potential}}, \bibinfo{journal}{Phys. Rev. B}
  \textbf{\bibinfo{volume}{93}}, \bibinfo{pages}{121303}
  (\bibinfo{year}{2016}).

\bibitem[{\citenamefont{Lagoudakis et~al.}(2010)\citenamefont{Lagoudakis,
  Pietka, Wouters, Andr\'e, and Deveaud-Pl\'edran}}]{Lagoudakis:PRL2010}
\bibinfo{author}{\bibfnamefont{K.~G.} \bibnamefont{Lagoudakis}},
  \bibinfo{author}{\bibfnamefont{B.}~\bibnamefont{Pietka}},
  \bibinfo{author}{\bibfnamefont{M.}~\bibnamefont{Wouters}},
  \bibinfo{author}{\bibfnamefont{R.}~\bibnamefont{Andr\'e}}, \bibnamefont{and}
  \bibinfo{author}{\bibfnamefont{B.}~\bibnamefont{Deveaud-Pl\'edran}},
  \emph{\bibinfo{title}{Coherent Oscillations in an Exciton-Polariton Josephson
  Junction}}, \bibinfo{journal}{Phys. Rev. Lett.}
  \textbf{\bibinfo{volume}{105}}, \bibinfo{pages}{120403}
  (\bibinfo{year}{2010}).

\bibitem[{\citenamefont{Veit et~al.}(2012)\citenamefont{Veit, A\ss{}mann,
  Bayer, L\"offler, H\"ofling, Kamp, and Forchel}}]{Veit:PRB2012}
\bibinfo{author}{\bibfnamefont{F.}~\bibnamefont{Veit}},
  \bibinfo{author}{\bibfnamefont{M.}~\bibnamefont{A\ss{}mann}},
  \bibinfo{author}{\bibfnamefont{M.}~\bibnamefont{Bayer}},
  \bibinfo{author}{\bibfnamefont{A.}~\bibnamefont{L\"offler}},
  \bibinfo{author}{\bibfnamefont{S.}~\bibnamefont{H\"ofling}},
  \bibinfo{author}{\bibfnamefont{M.}~\bibnamefont{Kamp}}, \bibnamefont{and}
  \bibinfo{author}{\bibfnamefont{A.}~\bibnamefont{Forchel}},
  \emph{\bibinfo{title}{Spatial dynamics of stepwise homogeneously pumped
  polariton condensates}}, \bibinfo{journal}{Phys. Rev. B}
  \textbf{\bibinfo{volume}{86}}, \bibinfo{pages}{195313}
  (\bibinfo{year}{2012}).

\bibitem[{\citenamefont{Ant\'on et~al.}(2013)\citenamefont{Ant\'on, Liew, Tosi,
  Mart\'{\i}n, Gao, Hatzopoulos, Eldridge, Savvidis, and
  Vi\~na}}]{Anton:PRB2013}
\bibinfo{author}{\bibfnamefont{C.}~\bibnamefont{Ant\'on}},
  \bibinfo{author}{\bibfnamefont{T.~C.~H.} \bibnamefont{Liew}},
  \bibinfo{author}{\bibfnamefont{G.}~\bibnamefont{Tosi}},
  \bibinfo{author}{\bibfnamefont{M.~D.} \bibnamefont{Mart\'{\i}n}},
  \bibinfo{author}{\bibfnamefont{T.}~\bibnamefont{Gao}},
  \bibinfo{author}{\bibfnamefont{Z.}~\bibnamefont{Hatzopoulos}},
  \bibinfo{author}{\bibfnamefont{P.~S.} \bibnamefont{Eldridge}},
  \bibinfo{author}{\bibfnamefont{P.~G.} \bibnamefont{Savvidis}},
  \bibnamefont{and} \bibinfo{author}{\bibfnamefont{L.}~\bibnamefont{Vi\~na}},
  \emph{\bibinfo{title}{Energy relaxation of exciton-polariton condensates in
  quasi-one-dimensional microcavities}}, \bibinfo{journal}{Phys. Rev. B}
  \textbf{\bibinfo{volume}{88}}, \bibinfo{pages}{035313}
  (\bibinfo{year}{2013}).

\bibitem[{\citenamefont{Ardizzone et~al.}(2012)\citenamefont{Ardizzone,
  Abbarchi, Lemaitre, Sagnes, Senellart, Bloch, Delalande, Tignon, and
  Roussignol}}]{Ardizzone:PRB2012}
\bibinfo{author}{\bibfnamefont{V.}~\bibnamefont{Ardizzone}},
  \bibinfo{author}{\bibfnamefont{M.}~\bibnamefont{Abbarchi}},
  \bibinfo{author}{\bibfnamefont{A.}~\bibnamefont{Lemaitre}},
  \bibinfo{author}{\bibfnamefont{I.}~\bibnamefont{Sagnes}},
  \bibinfo{author}{\bibfnamefont{P.}~\bibnamefont{Senellart}},
  \bibinfo{author}{\bibfnamefont{J.}~\bibnamefont{Bloch}},
  \bibinfo{author}{\bibfnamefont{C.}~\bibnamefont{Delalande}},
  \bibinfo{author}{\bibfnamefont{J.}~\bibnamefont{Tignon}}, \bibnamefont{and}
  \bibinfo{author}{\bibfnamefont{P.}~\bibnamefont{Roussignol}},
  \emph{\bibinfo{title}{Bunching visibility of optical parametric emission in a
  semiconductor microcavity}}, \bibinfo{journal}{Phys. Rev. B}
  \textbf{\bibinfo{volume}{86}}, \bibinfo{pages}{041301}
  (\bibinfo{year}{2012}).

\bibitem[{\citenamefont{Lecomte et~al.}(2014)\citenamefont{Lecomte, Taj,
  Lemaitre, Bloch, Delalande, Tignon, and Roussignol}}]{Lecomte:PRB2014}
\bibinfo{author}{\bibfnamefont{T.}~\bibnamefont{Lecomte}},
  \bibinfo{author}{\bibfnamefont{D.}~\bibnamefont{Taj}},
  \bibinfo{author}{\bibfnamefont{A.}~\bibnamefont{Lemaitre}},
  \bibinfo{author}{\bibfnamefont{J.}~\bibnamefont{Bloch}},
  \bibinfo{author}{\bibfnamefont{C.}~\bibnamefont{Delalande}},
  \bibinfo{author}{\bibfnamefont{J.}~\bibnamefont{Tignon}}, \bibnamefont{and}
  \bibinfo{author}{\bibfnamefont{P.}~\bibnamefont{Roussignol}},
  \emph{\bibinfo{title}{Polariton-polariton interaction potentials
  determination by pump-probe degenerate scattering in a multiple
  microcavity}}, \bibinfo{journal}{Phys. Rev. B} \textbf{\bibinfo{volume}{89}},
  \bibinfo{pages}{155308} (\bibinfo{year}{2014}).

\bibitem[{\citenamefont{Jacqmin et~al.}(2014)\citenamefont{Jacqmin, Carusotto,
  Sagnes, Abbarchi, Solnyshkov, Malpuech, Galopin, Lema\^{\i}tre, Bloch, and
  Amo}}]{Jacqmin:PRL2014}
\bibinfo{author}{\bibfnamefont{T.}~\bibnamefont{Jacqmin}},
  \bibinfo{author}{\bibfnamefont{I.}~\bibnamefont{Carusotto}},
  \bibinfo{author}{\bibfnamefont{I.}~\bibnamefont{Sagnes}},
  \bibinfo{author}{\bibfnamefont{M.}~\bibnamefont{Abbarchi}},
  \bibinfo{author}{\bibfnamefont{D.~D.} \bibnamefont{Solnyshkov}},
  \bibinfo{author}{\bibfnamefont{G.}~\bibnamefont{Malpuech}},
  \bibinfo{author}{\bibfnamefont{E.}~\bibnamefont{Galopin}},
  \bibinfo{author}{\bibfnamefont{A.}~\bibnamefont{Lema\^{\i}tre}},
  \bibinfo{author}{\bibfnamefont{J.}~\bibnamefont{Bloch}}, \bibnamefont{and}
  \bibinfo{author}{\bibfnamefont{A.}~\bibnamefont{Amo}},
  \emph{\bibinfo{title}{Direct Observation of Dirac Cones and a Flatband in a
  Honeycomb Lattice for Polaritons}}, \bibinfo{journal}{Phys. Rev. Lett.}
  \textbf{\bibinfo{volume}{112}}, \bibinfo{pages}{116402}
  (\bibinfo{year}{2014}).

\bibitem[{\citenamefont{Kokhanchik et~al.}(2020)\citenamefont{Kokhanchik,
  Sigurdsson, Pietka, Szczytko, and Lagoudakis}}]{Kokhanchik:arxive2020}
\bibinfo{author}{\bibfnamefont{P.}~\bibnamefont{Kokhanchik}},
  \bibinfo{author}{\bibfnamefont{H.}~\bibnamefont{Sigurdsson}},
  \bibinfo{author}{\bibfnamefont{B.}~\bibnamefont{Pietka}},
  \bibinfo{author}{\bibfnamefont{J.}~\bibnamefont{Szczytko}}, \bibnamefont{and}
  \bibinfo{author}{\bibfnamefont{P.~G.} \bibnamefont{Lagoudakis}},
  \emph{\bibinfo{title}{Photonic {Berry} curvature in double liquid crystal
  microcavities with broken inversion symmetry}}, \bibinfo{journal}{arXiv}
  (\bibinfo{year}{2020}), \eprint{2009.07189}.

\bibitem[{\citenamefont{Gao et~al.}(2012)\citenamefont{Gao, Eldridge, Liew,
  Tsintzos, Stavrinidis, Deligeorgis, Hatzopoulos, and Savvidis}}]{Gao:PRB2012}
\bibinfo{author}{\bibfnamefont{T.}~\bibnamefont{Gao}},
  \bibinfo{author}{\bibfnamefont{P.~S.} \bibnamefont{Eldridge}},
  \bibinfo{author}{\bibfnamefont{T.~C.~H.} \bibnamefont{Liew}},
  \bibinfo{author}{\bibfnamefont{S.~I.} \bibnamefont{Tsintzos}},
  \bibinfo{author}{\bibfnamefont{G.}~\bibnamefont{Stavrinidis}},
  \bibinfo{author}{\bibfnamefont{G.}~\bibnamefont{Deligeorgis}},
  \bibinfo{author}{\bibfnamefont{Z.}~\bibnamefont{Hatzopoulos}},
  \bibnamefont{and} \bibinfo{author}{\bibfnamefont{P.~G.}
  \bibnamefont{Savvidis}}, \emph{\bibinfo{title}{Polariton condensate
  transistor switch}}, \bibinfo{journal}{Phys. Rev. B}
  \textbf{\bibinfo{volume}{85}}, \bibinfo{pages}{235102}
  (\bibinfo{year}{2012}).

\bibitem[{\citenamefont{Dasbach et~al.}(2002)\citenamefont{Dasbach, Schwab,
  Bayer, Krizhanovskii, and Forchel}}]{Dasbach:PRB2002}
\bibinfo{author}{\bibfnamefont{G.}~\bibnamefont{Dasbach}},
  \bibinfo{author}{\bibfnamefont{M.}~\bibnamefont{Schwab}},
  \bibinfo{author}{\bibfnamefont{M.}~\bibnamefont{Bayer}},
  \bibinfo{author}{\bibfnamefont{D.}~\bibnamefont{Krizhanovskii}},
  \bibnamefont{and} \bibinfo{author}{\bibfnamefont{A.}~\bibnamefont{Forchel}},
  \emph{\bibinfo{title}{Tailoring the polariton dispersion by optical
  confinement: Access to a manifold of elastic polariton pair scattering
  channels}}, \bibinfo{journal}{Phys. Rev. B} \textbf{\bibinfo{volume}{66}},
  \bibinfo{pages}{201201} (\bibinfo{year}{2002}).

\bibitem[{\citenamefont{Ciuti}(2004)}]{Ciuti:PRB2004}
\bibinfo{author}{\bibfnamefont{C.}~\bibnamefont{Ciuti}},
  \emph{\bibinfo{title}{Branch-entangled polariton pairs in planar
  microcavities and photonic wires}}, \bibinfo{journal}{Phys. Rev. B}
  \textbf{\bibinfo{volume}{69}}, \bibinfo{pages}{245304}
  (\bibinfo{year}{2004}).

\bibitem[{\citenamefont{St-Jean et~al.}(2017)\citenamefont{St-Jean, Goblot,
  Galopin, Lema{\^\i}tre, Ozawa, Le~Gratiet, Sagnes, Bloch, and
  Amo}}]{StJean:NaturePhot2017}
\bibinfo{author}{\bibfnamefont{P.}~\bibnamefont{St-Jean}},
  \bibinfo{author}{\bibfnamefont{V.}~\bibnamefont{Goblot}},
  \bibinfo{author}{\bibfnamefont{E.}~\bibnamefont{Galopin}},
  \bibinfo{author}{\bibfnamefont{A.}~\bibnamefont{Lema{\^\i}tre}},
  \bibinfo{author}{\bibfnamefont{T.}~\bibnamefont{Ozawa}},
  \bibinfo{author}{\bibfnamefont{L.}~\bibnamefont{Le~Gratiet}},
  \bibinfo{author}{\bibfnamefont{I.}~\bibnamefont{Sagnes}},
  \bibinfo{author}{\bibfnamefont{J.}~\bibnamefont{Bloch}}, \bibnamefont{and}
  \bibinfo{author}{\bibfnamefont{A.}~\bibnamefont{Amo}},
  \emph{\bibinfo{title}{Lasing in topological edge states of a one-dimensional
  lattice}}, \bibinfo{journal}{Nat. Photonics} \textbf{\bibinfo{volume}{11}},
  \bibinfo{pages}{651} (\bibinfo{year}{2017}).

\bibitem[{\citenamefont{Yamamoto et~al.}(2017)\citenamefont{Yamamoto, Aihara,
  Leleu, Kawarabayashi, Kako, Fejer, Inoue, and
  Takesue}}]{Yamamoto:NPJQInf2017}
\bibinfo{author}{\bibfnamefont{Y.}~\bibnamefont{Yamamoto}},
  \bibinfo{author}{\bibfnamefont{K.}~\bibnamefont{Aihara}},
  \bibinfo{author}{\bibfnamefont{T.}~\bibnamefont{Leleu}},
  \bibinfo{author}{\bibfnamefont{K.-i.} \bibnamefont{Kawarabayashi}},
  \bibinfo{author}{\bibfnamefont{S.}~\bibnamefont{Kako}},
  \bibinfo{author}{\bibfnamefont{M.}~\bibnamefont{Fejer}},
  \bibinfo{author}{\bibfnamefont{K.}~\bibnamefont{Inoue}}, \bibnamefont{and}
  \bibinfo{author}{\bibfnamefont{H.}~\bibnamefont{Takesue}},
  \emph{\bibinfo{title}{{Coherent Ising machines—Optical} neural networks
  operating at the quantum limit}}, \bibinfo{journal}{NPJ Quantum Inf.}
  \textbf{\bibinfo{volume}{3}}, \bibinfo{pages}{1} (\bibinfo{year}{2017}).

\bibitem[{\citenamefont{Opala et~al.}(2018)\citenamefont{Opala, Pieczarka, and
  Matuszewski}}]{Opala:PRB2018}
\bibinfo{author}{\bibfnamefont{A.}~\bibnamefont{Opala}},
  \bibinfo{author}{\bibfnamefont{M.}~\bibnamefont{Pieczarka}},
  \bibnamefont{and}
  \bibinfo{author}{\bibfnamefont{M.}~\bibnamefont{Matuszewski}},
  \emph{\bibinfo{title}{Theory of relaxation oscillations in exciton-polariton
  condensates}}, \bibinfo{journal}{Phys. Rev. B} \textbf{\bibinfo{volume}{98}},
  \bibinfo{pages}{195312} (\bibinfo{year}{2018}).

\bibitem[{\citenamefont{Ballarini et~al.}(2020)\citenamefont{Ballarini,
  Gianfrate, Panico, Opala, Ghosh, Dominici, Ardizzone, De~Giorgi, Lerario,
  Gigli et~al.}}]{Ballarini:NanoLett2020}
\bibinfo{author}{\bibfnamefont{D.}~\bibnamefont{Ballarini}},
  \bibinfo{author}{\bibfnamefont{A.}~\bibnamefont{Gianfrate}},
  \bibinfo{author}{\bibfnamefont{R.}~\bibnamefont{Panico}},
  \bibinfo{author}{\bibfnamefont{A.}~\bibnamefont{Opala}},
  \bibinfo{author}{\bibfnamefont{S.}~\bibnamefont{Ghosh}},
  \bibinfo{author}{\bibfnamefont{L.}~\bibnamefont{Dominici}},
  \bibinfo{author}{\bibfnamefont{V.}~\bibnamefont{Ardizzone}},
  \bibinfo{author}{\bibfnamefont{M.}~\bibnamefont{De~Giorgi}},
  \bibinfo{author}{\bibfnamefont{G.}~\bibnamefont{Lerario}},
  \bibinfo{author}{\bibfnamefont{G.}~\bibnamefont{Gigli}},
  \bibinfo{author}{\bibfnamefont{T.~C.~H.} \bibnamefont{Liew}},
  \bibinfo{author}{\bibfnamefont{M.}~\bibnamefont{Matuszewski}},
  \bibnamefont{and} \bibinfo{author}{\bibfnamefont{D.}~\bibnamefont{Sanvitto}},
  \emph{\bibinfo{title}{Polaritonic Neuromorphic Computing Outperforms Linear
  Classifiers}}, \bibinfo{journal}{Nano Lett.} \textbf{\bibinfo{volume}{20}},
  \bibinfo{pages}{3506} (\bibinfo{year}{2020}).

\bibitem[{\citenamefont{Xue et~al.}(2021)\citenamefont{Xue, Chestnov, Sedov,
  Kiktenko, Fedorov, Schumacher, Ma, and Kavokin}}]{Xue:PRR2021}
\bibinfo{author}{\bibfnamefont{Y.}~\bibnamefont{Xue}},
  \bibinfo{author}{\bibfnamefont{I.}~\bibnamefont{Chestnov}},
  \bibinfo{author}{\bibfnamefont{E.}~\bibnamefont{Sedov}},
  \bibinfo{author}{\bibfnamefont{E.}~\bibnamefont{Kiktenko}},
  \bibinfo{author}{\bibfnamefont{A.~K.} \bibnamefont{Fedorov}},
  \bibinfo{author}{\bibfnamefont{S.}~\bibnamefont{Schumacher}},
  \bibinfo{author}{\bibfnamefont{X.}~\bibnamefont{Ma}}, \bibnamefont{and}
  \bibinfo{author}{\bibfnamefont{A.}~\bibnamefont{Kavokin}},
  \emph{\bibinfo{title}{Split-ring polariton condensates as macroscopic
  two-level quantum systems}}, \bibinfo{journal}{Phys. Rev. Research}
  \textbf{\bibinfo{volume}{3}}, \bibinfo{pages}{013099} (\bibinfo{year}{2021}).

\end{thebibliography}
\bibliographystyle{apsrev_my}

\section*{Data availability}
Data related to the figures can be found at https://doi.org/10.6084/m9.figshare.13713889.v1. Other data related to this work are available from the authors upon reasonable request.

\section*{Acknowledgments}
We acknowledge discussions with Alberto Amo and Jacek Kasprzak. We are grateful to the late Micha{\l} Nawrocki for his contribution to the initial stage of this work and late Jolanta Borysiuk for preparation of the sample for TEM measurements. This work was partially supported by the Polish National Science Center under decisions DEC-2013/10/E/ST3/00215, DEC-2017/25/N/ST3/00465 and DEC-2019/32/T/ST3/00332. The research was carried out with the use of CePT, CeZaMat and NLTK infrastructures financed by the European Union - the European Regional Development Fund within the Operational Programme ``Innovative economy'' for 2007-2013.

\end{document}